\documentclass[prx,aps,floatfix,10pt,twocolumn,nofootinbib,superscriptaddress]{revtex4-1}

\usepackage[T1]{fontenc}
\usepackage[utf8]{inputenc}
\usepackage{microtype}
\usepackage{array}
\usepackage{graphicx}
\usepackage{dcolumn}
\usepackage{bm}
\usepackage{amsmath}
\usepackage{amsfonts}
\usepackage{amssymb}
\usepackage{amstext}   
\usepackage{amsthm}
\usepackage{array}
\usepackage[english]{babel}
\usepackage{makecell}
\usepackage{dsfont}
\usepackage{nameref}
\usepackage{color}
\usepackage{float}
\usepackage[outdir=./]{epstopdf}
\usepackage{subfloat}
\usepackage[colorlinks = true, citecolor = blue, urlcolor =cyan]{hyperref}
\usepackage[figure]{hypcap}
\usepackage{multirow}
\usepackage{makecell}
\usepackage{xcolor}
\usepackage{mathtools}
\usepackage{mathdots}
\usepackage[notransparent]{svg}
\usepackage{layouts}
\usepackage{todonotes} 
\newcolumntype{C}{>{$}c<{$}}
\AtBeginDocument{
\heavyrulewidth=.08em
\lightrulewidth=.05em
\cmidrulewidth=.03em
\belowrulesep=.65ex
\belowbottomsep=0pt
\aboverulesep=.4ex
\abovetopsep=0pt
\cmidrulesep=\doublerulesep
\cmidrulekern=.5em
\defaultaddspace=.5em
\tabcolsep=7pt
}
\usepackage{booktabs}

\definecolor{emerald}{rgb}{0.07, 0.53, 0.03}

\usepackage{braket}

\usepackage{nicefrac} 

\usepackage[normalem]{ulem} 

\usepackage{soul}   
\setstcolor{red} 
\setul{}{1.5pt}  

\usepackage{paralist}

\begin{document}

\title{Autonomous Stabilization of Floquet States Using Static Dissipation}
\author{Martin Ritter}
\affiliation{Joint Quantum Institute, Department of Physics, University of Maryland, College Park, Maryland 20742, USA}
\author{David M. Long}
\affiliation{Joint Quantum Institute, Department of Physics, University of Maryland, College Park, Maryland 20742, USA}
\affiliation{Condensed Matter Theory Center, Department of Physics, University of Maryland, College Park, Maryland 20742, USA}
\affiliation{Department of Physics, Stanford University, Stanford, California 94305, USA}
\author{Qianao Yue}
\affiliation{Joint Quantum Institute, Department of Physics, University of Maryland, College Park, Maryland 20742, USA}
\author{Anushya Chandran}
\affiliation{Department of Physics, Boston University, Boston, Massachusetts 02215, USA}
\affiliation{Max-Planck-Institut f\"{u}r Physik komplexer Systeme, N\"othnitzer Str. 38, 01187 Dresden, Germany}
\author{Alicia J.  Koll\'ar}
\email{akollar@umd.edu}
\affiliation{Joint Quantum Institute, Department of Physics, University of Maryland, College Park, Maryland 20742, USA}

\preprint{APS/123-QED}

\date{\today}

\begin{abstract}
    Floquet engineering, in which the properties of a quantum system are modified through the application of strong periodic drives, is an indispensable tool in atomic and condensed matter systems. 
    However, it is inevitably limited by intrinsic heating processes.
    We describe a simple autonomous scheme, which exploits a \emph{static} coupling between the driven system and a lossy auxiliary, to cool large classes of Floquet systems into desired states.
    We present experimental and theoretical evidence for the stabilization of a chosen \emph{quasienergy} state in a strongly modulated transmon qubit coupled to an auxiliary microwave cavity with fixed frequency and photon loss.
    The scheme naturally extends to Floquet systems with multiple degrees of freedom.
    As an example, we demonstrate the stabilization of topological photon pumping in a driven cavity-QED system numerically. 
    The coupling to the auxiliary cavity increases the average photon current and the fidelity of non-classical states, such as high photon number Fock states, that can be prepared in the system cavity.
\end{abstract}

\maketitle

\section{Introduction}\label{sec:intro}
The ability to experimentally control and manipulate quantum systems has rapidly advanced over the last two decades. One indispensable tool in the quantum mechanic's toolkit is Floquet engineering, in which the properties of the system are qualitatively modified by strong periodic driving.
It underlies the creation of optical lattices and synthetic band structures~\cite{Ultracoldgases_review_Bloch, Dalibard_Floquet_review, Bukov2015, Oka:2009_graphene, Floquet_topological_insulators,Holthaus_2016}, artificial gauge fields~\cite{SpielmanRamanGage,Lin:2009, Dalibard_Floquet_review, RoushanChiralCurrent, rosen:2024_implementingsyntheticmagneticvector}, topological phases impossible in static systems~\cite{Rudner2013,Harper:2020aa, Wintersperger2020,khemani2019dtcreview, Else:2020_dtcreview}, topological charge and photon pumps~\cite{Thouless,Citro:2023_thoulesspumpreview,Martin:2017aa,Crowley:2019_classification,Nathan:2019_drivendiss,PSAROUDAKI2021,Long:2022boost,Long2021:class,Nathan2020b}, and exotic quantum codes~\cite{Hastings:2021_dynamically}, to name a few.

However, Floquet engineering is always limited by heating, or more correctly entropy production.
There are three conceptually distinct mechanisms for heating. 
Consider a single degree of freedom with a few energy levels subject to Floquet driving.
The stationary states of the isolated system (known as quasienergy states) involve admixtures of \emph{both} ground and excited states. 
Accounting for dissipation and dephasing in the laboratory frame leads to \emph{both} excitation and de-excitation processes in the quasienergy state basis, and thus mixed steady states with larger entropy than the stationary states of the isolated system.
Second, restricting to a few levels is always an approximation in a physical system. 
Higher order processes often allow the system to absorb energy from the drive, obtain weight on states outside of the intended operating subspace and thus heat up.
Third, with many interacting degrees of freedom, the system can absorb energy from the drive and locally become more mixed, even if the first two mechanisms are suppressed~\cite{Lazarides:2014yg, Ponte:2015zr,Mori:2018_prethermalreview}. 

We present a simple and general purpose dissipative stabilization scheme to mitigate the effects of the first type of (excited-state admixture) heating.
Fortuitously, our scheme addresses the other two types of heating as well, at least within certain physical contexts.

Conventional dissipative preparation methods leverage native decay processes to successfully cool into ground states~\cite{Happer:1972_optical_pumping, Phillips_cooling_nobel} or dark states~\cite{Fleishhauer_EIT_rev_mod_phys}.
However, in the presence of strong drives, laboratory frame decay processes cause \emph{both} excitation and de-excitation. This, combined with the unboundedness of Floquet spectra from above and below, causes conventional autonomous cooling schemes like optical pumping~\cite{Happer:1972_optical_pumping} 
and dark-state engineering~\cite{Diehl:2008, Fleishhauer_EIT_rev_mod_phys, Harrington:2022aa} to fail. Their generalization to the time-dependent case is not known.

One route to stabilizing more general, but still time-independent, states is to engineer new dissipation terms distinct from the native dissipators~\cite{Poyatos:1996aa, Harrington:2022aa}.
For example, in superconducting qubit architectures, two photon loss has been used to 
stabilize bosonic Kerr-Cat qubits~\cite{Leghtas:2015, Lescanne_Cat_qubit_2020, Putterman_Cat_qubit_2022}
, while selective photon addition can stabilize code spaces in a Cat code~\cite{Gertler:2021}. Collective dissipators can also stabilize target entangled states of atoms~\cite{Plenio:1999,Kraus:2008, Verstraete:2009, Barrerio:2011, Lin:2013, Shankar:2013, Brown:2022}. In principle, periodic modulation of engineered dissipators could implement spontaneous emission into a specific quasienergy state.
Such an approach is however highly fine-tuned, requiring knowledge of the target quasienergy state at all times during the drive cycle. 

A different route to stabilization is to actively extract heat from the system continuously~\cite{Home:2009_completeIonSet,Home:2009_sympatheticIon, Raghunandan:2020, Petiziol:2022aa, Wang:2024_0pi,Murch:2012,Lu:2017,Ma:2019ba,Li2024autonomous} or stroboscopically~\cite{Lloyd:1996aa, Sorovar:2005_resetCooling, Mi:2024} into an auxiliary system.
This type of approach has been used to stabilize arbitrary static single-qubit states~\cite{Murch:2012,Lu:2017,Ma:2019ba} and entangled states of two qubits~\cite{Shankar:2013,Li2024autonomous}. 
These methods have also been extended to Floquet systems in the limit of high-frequency driving, where they stabilize the ground state space of static effective Hamiltonians. Refs.~\cite{Petiziol:2022aa,Wang:2024_0pi} numerically demonstrated such stabilization, while Ref.~\cite{Mi:2024} experimentally cooled into the ground state of an effective many-qubit Hamiltonian by cycling between a Trotterized Hamiltonian evolution step and an auxiliary qubit reset step.

\begin{figure}[t]
    \centering
    \includegraphics[width=\linewidth]{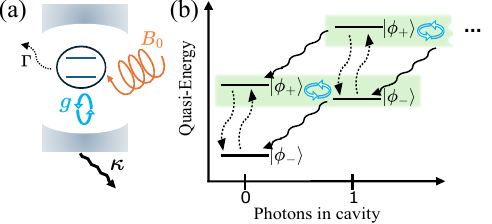}
    \caption{\textbf{Static dissipation stabilizes Floquet states.} 
    (a) Schematic of the augmented system consisting of a spin with intrinsic dissipation rate $\Gamma$ (dotted wavy arrow) subject to a strong circularly polarized drive of strength $B_0$ (orange), coupled to an auxiliary cavity with coupling strength $g$ (light blue) and photonic decay rate $\kappa$ (solid wavy arrow).
    (b)~Quasienergy level structure of the augmented system. The representative quasienergy states, $\ket{\phi_\pm}$, are shown for $n=0,1$ photons in the cavity. Far from resonance, photon decay drives the system to $n=0$ while preserving the spin state. On resonance however, the cavity coupling $g$ hybridizes spin states with different photon numbers, allowing the state $\ket{\phi_+}$ in the $n=0$ sector to convert to $\ket{\phi_-}$ and decay. When the spin decay rate $\Gamma$ is much weaker than $\kappa$, the unique steady state remains close to $\ket{\phi_-,n=0}$. 
    }
    \label{fig:intuition_cartoon}
\end{figure}


Here we show that, regardless of drive-frequency regime, a static coupling to an unmodulated but lossy auxiliary system can cool a Floquet system into an isolated target quasienergy state. This method is autonomous, requiring no active feedback, and exploits no microscopic knowledge of the time-dependence of the target quasienergy state. It instead introduces static loss to spectral regions with unwanted states. As a result, the method applies to a broad class of time-dependent Hamiltonians, well beyond the particular examples considered in this article, as long as the target state is not in a dense level spectrum. In particular, the method does not rely either on exact resonance of quasienergies of the non-target states, nor on any resonances in the instantaneous spectrum.

We present experimental and numerical evidence for stabilization in a strongly driven two-level system coupled to a lossy bosonic mode. The experimental system consists of a strongly modulated transmon qubit in the low-frequency regime coupled to an auxiliary microwave cavity with photon loss [Fig.~\ref{fig:intuition_cartoon}(a)]. Numerical simulations treat the combined system within the master equation framework and demonstrate stabilization in all drive regimes.

As an application beyond two-level systems, we develop the theory of a topological photon pump~\cite{Martin:2017aa,Crowley:2019_classification,Nathan:2019_drivendiss,PSAROUDAKI2021,Long:2022boost} coupled to an auxiliary lossy cavity. The topological photon pump is a strongly driven cavity-QED system with a long-lived photon current of topological origin. We show that the auxiliary cavity brings the value of the photon current closer to its ideal value, and increases the fidelity of high-photon number Fock states that are prepared in the cavity at special times~\cite{Long:2022boost}.

Unlike dissipation in the Floquet system, dissipation in the auxiliary does not result in heating.
In the limit of weak driving, the states of the Floquet system are near-degenerate and couple to the auxiliary symmetrically.
However, in the limit where the drive in the Floquet model is strong compared to the coupling to the auxiliary, hybridization between the Floquet states and the auxiliary is dependent on the quasienergy [Fig.~\ref{fig:intuition_cartoon}(b)].
When a resonance condition is satisfied, the loss in the auxiliary leads to the state $ \ket{d} = \ket{\mathrm{target}(t)}\otimes \ket{\mathrm{vac}}$ being an approximate attractor of the combined dynamics, with the Floquet system in the target quasienergy state $\ket{\mathrm{target}(t)}$ and the auxiliary in its vacuum state.
Dissipation and decoherence in the Floquet system drive transitions away from $\ket{d}$, but the loss in the auxiliary pumps the combined system back to $\ket{d}$. As long as the auxiliary is the strongest dissipation channel, the steady state of the primary system has a high degree of purity and significant overlap with $\ket{\mathrm{target}(t)}$. 

The remainder of the manuscript is organized as follows.
In Sec.~\ref{sec:model}, we describe the physical setup involving a periodically modulated spin coupled to a lossy cavity, which showcases the autonomous stabilization effect. 
In Sec.~\ref{sec:expt}, we present experimental evidence for stabilization in this physical setup in the low-frequency (adiabatic) regime.
We turn to the general theory in Sec.~\ref{sec:beyondAdiabatic}, and provide numerical evidence for stabilization in all frequency regimes. Lastly, in Sec.~\ref{sec:pumping}, we show that dissipative stabilization enhances the performance of the topological photon pump~\cite{Nathan:2019_drivendiss,Long:2022boost}.
We conclude in Sec.~\ref{sec:conclusion}, where we discuss further applications beyond two-level systems.

\section{The Physical Set-up}\label{sec:model}

Consider a minimal Floquet model of a single spin in a rotating magnetic field. The Hamiltonian is,
\begin{equation}\label{eqn:Hmodel}
    H_B = \frac{1}{2}\vec{\sigma}\cdot\vec{B}(t) = \frac{B_0}{2}\cos(\omega_{\mathrm{mod}} t) \sigma_z + \frac{B_0}{2}\sin(\omega_{\mathrm{mod}} t) \sigma_x,
\end{equation}
with a fixed modulation frequency $\omega_{\mathrm{mod}}$ and constant magnetic field amplitude $B_0$. The model happens to be soluble (it reduces to a static problem in the frame rotating about the $y$-axis with frequency $\omega_{\mathrm{mod}}$). This solubility is however irrelevant to what follows, as the augmented system with dissipation is not soluble. See the schematic in Fig.~\ref{fig:intuition_cartoon}(a). 

The quasienergy states, or Floquet states, are special solutions to the Schr\"odinger equation with the same periodicity as the drive. They take the form
\begin{equation}
\label{Eq:QEStateDef}
    \ket{\psi_\pm(t)} = e^{-i \epsilon_\pm t} \ket{\phi_\pm(t)},
\end{equation}
where ``\(\pm\)'' label the two states of the qubit, \(\ket{\phi_\pm(t)} = \ket{\phi_\pm(t+T_{\mathrm{mod}})}\) is the periodic quasienergy state, and \(\epsilon_\pm\) is its corresponding quasienergy. The quasienergy states are not unique: the same physical solutions \(\ket{\psi_\pm(t)}\) are obtained if we replace \(\epsilon_\pm \to \epsilon_\pm + m \omega_{\mathrm{mod}} = \epsilon_\pm^{(m)}\) and \(\ket{\phi_\pm(t)} \to e^{i m \omega_{\mathrm{mod}}t} \ket{\phi_\pm(t)} = \ket{\phi_\pm^{(m)}(t)}\) for $m$ an integer. Thus, we obtain countably many equivalent copies of the quasienergy states \(\ket{\phi_\pm^{(m)}(t)}\) with quasieneries differing by integer multiples of the drive (angular) frequency \(\omega_{\mathrm{mod}}\).

As is the usual convention, we select \(\ket{\phi_\pm^{(0)}(t)}\) from among these countably many choices by demanding \(|\epsilon^{(0)}_\pm| < \omega_{\mathrm{mod}}/2\). [For the model in Eq.~\eqref{eqn:Hmodel}, this choice imposes $\epsilon_+^{(0)} = -\epsilon_-^{(0)}$.] The interval of quasienergies \(\epsilon \in [(m-\tfrac{1}{2})\omega_{\mathrm{mod}}, (m+\tfrac{1}{2})\omega_{\mathrm{mod}})\) is called the \(m\)th Floquet zone, in analogy to Brillouin zones of quasimomentum. In the intermediate and high frequency regimes, we work with the states \(\ket{\phi_\pm^{(0)}(t)}\) belonging to the zeroth zone. In the adiabatic regime of small \(\omega_{\mathrm{mod}}\) however, it is helpful to use representative quasienergy states belonging to different zones (see the discussion at the beginning of Sec.~\ref{sec:expt}). In all cases, we refer to the chosen two representative quasienergy states as $\ket{\phi_\pm}$, without the Floquet index and with time-dependence implicit, and to their quasi-energies as $\epsilon_\pm$.

Static dissipation on the spin [denoted by $\Gamma$ in Fig.~\ref{fig:intuition_cartoon}(a)] generically leads to a mixed steady state, even if the spin is initialized in a quasienergy state of $H_B$. Depending on the time $t$ within the period, the static dissipation process primarily excites or de-excites the spin in the quasienergy basis; as both processes can happen, the steady state is mixed.

Our main result is that an additional \emph{static} coupling to an auxiliary system with loss can lead to steady states which are pure and which have high overlap with a chosen quasienergy state of $H_B$. For specificity, the auxiliary is taken to be a lossy bosonic mode, shown schematically as an optical cavity with photonic dissipation rate $\kappa$ in Fig.~\ref{fig:intuition_cartoon}(a). However, the result holds for more general auxiliary systems that have a characteristic energy scale and relax to a unique state, such as a lossy qubit, or a collection of bosonic modes in an energy band.

The augmented spin-cavity system is governed by the Hamiltonian
\begin{equation}\label{eqn:experimentH}
H_{B}^+ = \frac{1}{2}\vec{\sigma}\cdot\vec{B}(t) + \Delta\, a^\dagger a + g\,(a^\dagger \sigma^- + a \sigma^+),
\end{equation}
where $\Delta$ denotes the effective cavity frequency and $g$ denotes the strength of the spin-cavity coupling. 
The Hilbert space corresponding to $H_B^+$ is spanned by the states $\ket{\phi_\pm, n}$, where $n$ is the number of photons in the cavity. The pairs of quasienergy states for $n=0$ and $n=1$ are shown in Fig.~\ref{fig:intuition_cartoon}(b) along with the couplings induced by the coherent light-matter interaction (blue solid arrows) and the photonic dissipation (black wavy arrows). The hierarchy of energy scales that we assume is:
\begin{align}
\label{Eq:HierarchyScales}
    \Gamma < \kappa \lesssim g \ll \Delta, |\epsilon_+ - \epsilon_-|.
\end{align}
To effectively cool the system to a pure state, the cooling rate of the system $\kappa$ must exceed the other decay rates in the system ($\Gamma$) which can act as heating rates in the quasi-energy basis.
While a large $\kappa$ is desirable for faster cooling, exceeding the qubit-cavity coupling strength ($\kappa>g$) leads to a decoupling of the cavity from the main system, analogous to the bad cavity limit in cavity QED systems or the quantum Zeno effect \cite{Misra_Quantum_Zeno_theory_1977,Itano_Quantum_zeno_exp_1990}.
While the drive frequency \(\omega_\mathrm{mod}\) does not enter the hierarchy in Eq.~\eqref{Eq:HierarchyScales} explicitly, the quasienergies \(\epsilon_{\pm}\) depend on \(\omega_\mathrm{mod}\). Thus, the scale of \(\omega_\mathrm{mod}\) is indirectly constrained in terms of the other parameters by requiring that the quasienergy difference \(|\epsilon_+ - \epsilon_-|\) be large.
The hierarchy in Eq.~\eqref{Eq:HierarchyScales} is realized when the driving is strong, and the cavity-QED coupling is in the intermediate to strong coupling regime. The representative quasienergies in Eq. \eqref{Eq:HierarchyScales} depend on the drive frequency regime.

Stabilization is a consequence of resonances between states in the augmented system. Dissipative processes in the spin lead to transitions between $\ket{\phi_\pm, n}$, while photon loss reduces the photon number without affecting the spin state, relaxing the system to $\ket{\phi_\pm, 0}$. However, if the cavity frequency $\Delta$ is comparable to the quasienergy difference between $\ket{\phi_\pm}$, then there is a resonance between the $\ket{\phi_+, n-1}$ and $\ket{\phi_-,n}$ states. 
In the hierarchy of scales in Eq.~\eqref{Eq:HierarchyScales}, the spin-cavity coupling $g$ is the strongest coupling between these two resonant states. 
As shown in Fig.~\ref{fig:intuition_cartoon}(b), it causes the two states to hybridize, opening up a pathway for $\ket{\phi_+, 0}$ to decay to $\ket{\phi_-, 0}$. 
As there is no $\ket{\phi_+, -1}$ state, the state $\ket{\phi_-, 0}$ becomes an attractor of the photonic dissipation. 
The much weaker spin dissipation rate $\Gamma$ simply perturbs this attractor.
The augmented spin-cavity system thus achieves autonomous stabilization to an approximate product state, in which the spin is time-dependent and in a single quasienergy state $|\phi_-(t)\rangle$, while the cavity is close to empty.

\section{Experimental Observation of Stabilization in the Adiabatic Regime}\label{sec:expt}
In this section, we present an experimental implementation of the model in Eq.~\eqref{eqn:experimentH} in the adiabatic limit (where $\omega_{\mathrm{mod}}\ll B_0$) and demonstrate stabilization of a quasienergy state using a transmon qubit \cite{Blais_2021, Krantz:2019} coupled to a superconducting cavity. We show that, when the resonance condition in Fig.~\ref{fig:intuition_cartoon}(b) holds, the transmon is stabilized to a chosen instantaneous eigenstate for timescales that far exceed the native $T_1$ and $T_2$ decoherence times of the qubit. 

\begin{figure}[t]
\centering
    \includegraphics[width=\linewidth]{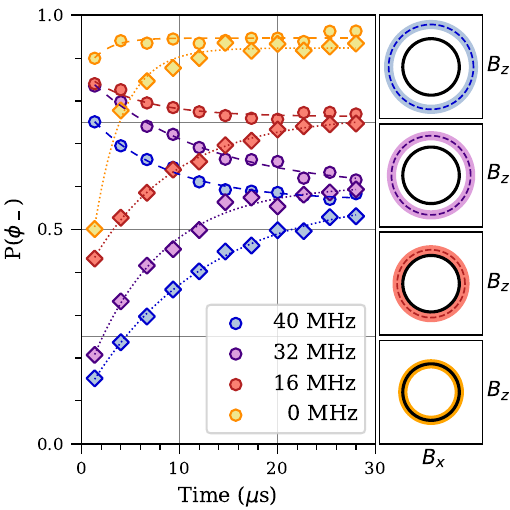}
    \caption{\label{fig:adiabatic_detuning_scan} 
    \textbf{Onset of stabilization in the quasienergy basis in the adiabatic limit.} 
    Measurements of the $\ket{\phi_-}$ population (averaged over two periods) as a function of evolution time. Diamond (circle) markers denote $\ket{\phi_+}$ ($\ket{\phi_-}$) initial state preparation for four different detuning values from the cavity. The error bars are smaller than the markers (see appendices \ref{appendix:B_field_synthesis} and \ref{appendix:error_bars} for measurement details).
    Curves are labeled by the detuning $\delta /2 \pi = (\Delta - B_0)/2\pi$. Side panels show the applied $2 \pi \times 80$~MHz rotating magnetic field (solid line) and the $\delta = 0$ ring for each detuning (dashed line). The shaded ring indicates the region of $\pm g = \pm 2\pi \times 13$~MHz around cavity resonance where the qubit states can hybridize strongly with the cavity. As $\delta \rightarrow 0$, the system steady state approaches the pure state $\ket{\phi_-}$ with increasing fidelity. 
    Exponential fits to the data are shown in dashed (dotted) lines for preparation in $\ket{\phi_-}$ ($\ket{\phi_+}$), with decay times ranging from 12~$\mu$s in the far-detuned regime, to 3~$\mu$s on resonance, consistent with the expected resonant decay rate of roughly \(2/\kappa = 3.8\,\mu\)s.
    The full detuning dependence of the stabilization time can be found in Appendix~\ref{appendix:timescales_and_contrast}.
    }
\end{figure}

\begin{figure*}[t]
	\begin{center}
		\includegraphics[width=\textwidth]{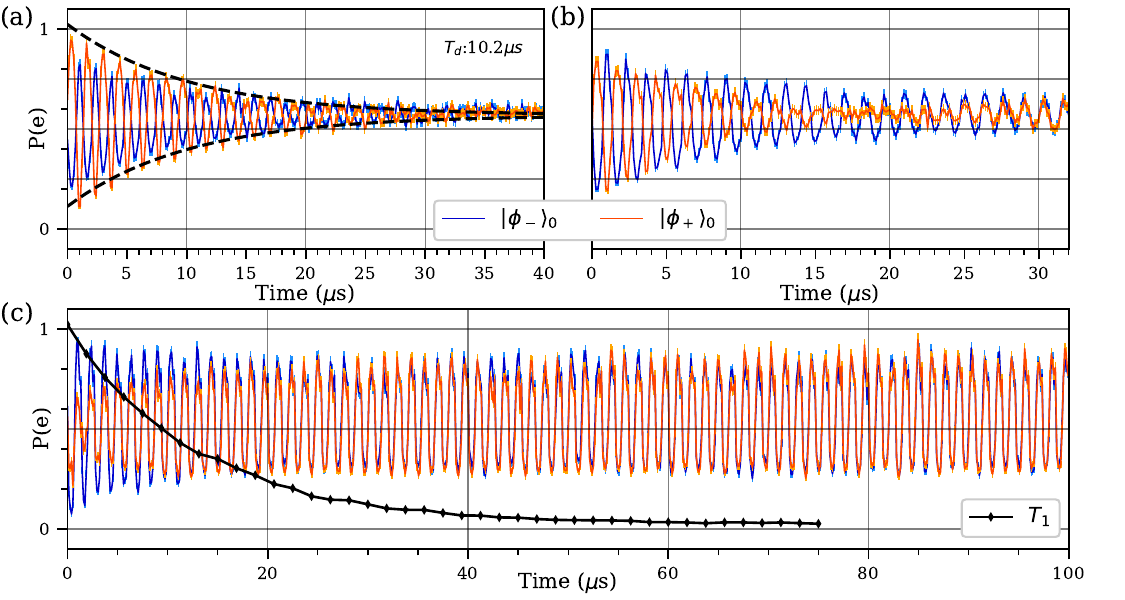}
	\end{center}
	\vspace{-0.6cm}
	\caption{\label{fig:hourglass} 
    \textbf{Dissipative stabilization of a quasienergy state in the adiabatic limit. }
    The panels show the probability of being in the $e$ or $\ket{\uparrow}$ state (obtained from fast ramp-out measurements, see Appendix~\ref{appendix:schedule} for details) after evolution under the Hamiltonian $H_B^+$ with $B_0 = 2\pi \times 80$~MHz and $\omega_{\mathrm{mod}} = 2\pi \times 0.75$~MHz for three different cavity detunings. 
    The orange (blue) curve is obtained by starting with the qubit aligned (anti-aligned) with the field, which we denote as $\ket{\phi_+}_0$ and $\ket{\phi_-}_0$, respectively. When there is perfect adiabatic following (and no photon or qubit decay), these two curves are sinusoidal with unit amplitude and are exactly out-of-phase.
    (a)~Far from resonance ($\delta = 6.15 g = 2\pi \times 80$~MHz), the curves oscillate out-of-phase for all times, but with a decaying amplitude. At long times, both states are close to a 50-50 mixture in the $z$-basis. Dashed line indicates a fit to the $\ket{\phi_+}_0$ data with extracted time constant $T_d = 10.2\  \mu$s. 
    (b) At intermediate detuning $\delta = 2.31g = 2\pi \times 30$~MHz, the blue trace shows persistent oscillations, while the orange trace first decays to zero, and re-emerges as a curve with the same amplitude and phase as the blue curve. As the steady state contrast is not unity, the stabilized state is a mixture with greater weight on the $|\phi_-\rangle$ state.
    (c) Closer to resonance, $\delta = 0.77g = 2\pi \times 10$~MHz, the blue curve is the attractor with a high contrast in the steady state, indicating that $|\phi_-\rangle$ is stabilized with high fidelity.
    The steady-state oscillations persist for far longer than any of the system coherence times, for example, the qubit $T_1$ decay shown for reference in black. 
    For details on error bar computation, see Appendix~\ref{appendix:error_bars}.
}
\end{figure*}


In the adiabatic limit, the quasienergy states are close to instantaneous eigenstates. The representatives $\ket{\phi_\pm}$ are the states most strongly coupled by the cavity and are several Floquet zones apart; indeed, their quasienergy splitting is close to the instantaneous splitting, $|\epsilon_+ - \epsilon_-| \approx B_0$. From Fig.~\ref{fig:intuition_cartoon}(b), the resonance condition is then:
\begin{align}
\label{Eq:AdLimitResonance}
\delta = \Delta - B_0 \approx 0.
\end{align}
We refer to the representative quasienergy states and the instantaneous eigenstates of $H_B$ as $\ket{\phi_\pm}$ below.

To create an \emph{effective} rotating magnetic field of the form in $H_B$ [Eq.~\eqref{eqn:Hmodel}], we combine simultaneous frequency modulation of the $g-e$ transition of the transmon and strong microwave driving at the mean $g-e$ transition frequency \cite{Long:2022boost}.
The flux bias sets the frequency of the $g-e$ transition \cite{Blais_2021, Krantz:2019}, and therefore creates an effective $B_z$. 
On the other hand, the  Rabi rate induced by a \emph{constant} AC microwave drive $A \cos(\omega_{d} t)$ implements an effective \emph{static} $B_x$ term in a frame rotating at the drive frequency $\omega_d$ \cite{wallsMilburn}.
To synthesize $H_B$, we first select an operating point using a DC flux bias $\Phi_{\mathrm{ext}} = \Phi_\mathrm{static}$, then set the microwave drive frequency to be resonant with the qubit transition at this bias point, and operate in a rotating frame at this frequency.
Rotation of the effective field at frequency $\omega_{\mathrm{mod}}$ is induced by modulating \emph{both} the flux bias $[\Phi_{\mathrm{ext}} = \Phi_\mathrm{static} + \delta \Phi \sin(\omega_{\mathrm{mod}}t)]$ and the AC drive $[V_{\mathrm{drive}} = A\cos(\omega_{\mathrm{mod}} t) \cos(\omega_d t)]$ in quadrature. See Appendix \ref{appendix:B_field_synthesis} for more details on the effective rotating field. 
In order to be deep within the adiabatic regime, we use $B_0 = 2\pi \times 80$~MHz, and modulation frequencies of $\omega_{\mathrm{mod}} = 2\pi\times0.75$~MHz. 

The auxiliary system is a superconducting resonator with effective frequency $\Delta=2\pi \times 40 - 140$~MHz, linewidth $\kappa = 2\pi\times 84$~kHz, and coupling strength $g = 2 \pi \times 13$~MHz to the transmon. 
The effective frequency $\Delta$ is the detuning between the cavity and the \emph{DC frequency set point} of the tunable qubit, and is varied using the latter.
The coupling strength $g$ satisfies Eq.~\eqref{Eq:HierarchyScales}, but is over two orders of magnitude larger than $\kappa$. 
As the time-scale for stabilization is controlled by the photon lifetime $\kappa^{-1}$, this allows for a larger time window within which to observe the decay from non-stabilized states. 

In principle, decay and dephasing processes in the qubit both contribute to mixing in the Floquet basis. (See Appendix~\ref{appendix:qubit_timescales} for qubit characterization details.)
However, the presence of the constant strong drive suppresses the effect of $T_2$ fluctuations in the qubit frequency, and the qubit follows the drive for $T_d \gg T_2$, even in the absence of stabilization from the cavity.  
We measure $T_d = 10.2 \ \mu$s empirically from the decay of adiabatic following at large cavity detuning, $\delta = 2\pi \times 80$~MHz shown in Fig.~\ref{fig:hourglass}(a).
As the lifetime of the cavity $1/\kappa = 1.9\ \mu$s is smaller than $T_d$, we have the required hierarchy of dissipative scales in Eq.~\eqref{Eq:HierarchyScales}.

In the adiabatic regime, the resonance condition is alternatively visualized in the $B_x-B_z$ plane. The trajectory traced out by $\vec{B}(t)$ on the plane is a closed curve (black curves, right panels of Fig.~\ref{fig:adiabatic_detuning_scan}), while the resonance condition $\delta = \Delta - |\vec{B}(t)| = 0$ defines a circle of radius $\Delta$ (dotted, right panels of Fig.~\ref{fig:adiabatic_detuning_scan}). Dissipative stabilization occurs in the vicinity of every intersection point of the two curves. In the simple model in Eq.~\eqref{eqn:Hmodel}, stabilization thus occurs at all times during the protocol, provided \(\Delta \approx B_0\). 
However, this is unnecessary; as long as the trajectory traced out by $\vec{B}(t)$ spends a significant fraction of time within roughly $g$ of the $\delta=0$ ring, photonic dissipation stabilizes the chosen quasienergy state, as discussed in Appendix~\ref{appendix:detuning_sweep_bare_basis}.
The stabilized cavity-state boosting protocol presented in Sec.~\ref{sec:pumping} uses this property in an essential manner.

The dynamics of the augmented qubit-cavity system are most transparent if the initial state of the spin coincides with a quasienergy state. In the adiabatic regime, this is an instantaneous eigenstate of $H_B$ at the start of the protocol.
Since generating $H_B$ requires strong driving fields, pulsed preparation of the instantaneous eigenstates is challenging, and we instead make use of an adiabatic preparation scheme. 
An optional $\pi$-pulse initializes the qubit in the bare $\ket{g}$ or $\ket{e}$ states, which are then mapped to $\ket{\phi_+}$ and $\ket{\phi_-}$, respectively, by the adiabatic preparation step.
After evolution under $H_B^+$ for a variable amount of time, the final state is mapped back to the bare transmon basis using either (i) a slow ramp-out protocol in which the field is rotated to a fixed point on the Bloch sphere to map the adiabatic states to the bare qubit states before measurement, or (ii) a fast ramp-out protocol in which the fields are shut off as quickly as possible to preserve the instantaneous qubit populations in the $\sigma_z$ basis. See Appendix~\ref{appendix:schedule} for details of the full waveforms and the different ramp out protocols.

The slow ramp-out measurements, such as those shown in Fig.~\ref{fig:adiabatic_detuning_scan}, approximately measure the population in the instantaneous eigenbasis of the system: $\ket{\phi_-}, \ket{\phi_+}$. In this basis, the measurement axis rotates along with the applied field, removing the time dependence of the rotating field. Far from resonance, both initial states ($\ket{\phi_-}_0$ and $\ket{\phi_+}_0$) undergo decay to a mixed state with $P(\phi_-)\approx0.5$, albeit with a small bias due to imperfections in the adiabatic ramp out. Strikingly, the rate at which the system decoheres is not limited by the native $T_2$ phase coherence time (approximately $0.5\,\mu$s) but instead is much closer to the $T_1$ time of $13\,\mu$s. 

Near resonance ($|\delta|\leq g$), the states $\ket{\phi_+,0}$ and $\ket{\phi_-,1}$ hybridize leading to preferential loss for $\ket{\phi_+}$ states through the cavity. This loss asymmetry shifts the steady-state of the system from a mixed state to an increasingly pure state, $\ket{\phi_-,0}$, as the detuning approaches zero [yellow curves in Fig.~\ref{fig:adiabatic_detuning_scan}]. 
In this regime, if the qubit is initialized in $\ket{\phi_-}_0$, it will stay in that state indefinitely whereas a qubit initialized in $\ket{\phi_+}_0$ will be dissipatively driven to the $\ket{\phi_-}$ state. In particular, the rate at which the system reaches steady state increases as the detuning approaches zero leading to a faster and stronger stabilization effect on resonance. 

As the detuning increases, the steady state fidelity $P(\phi_-)$ decreases, while the steady state relaxation time increases (see Appendix~\ref{appendix:timescales_and_contrast} for full detuning dependence). The inverse correlation between the steady state fidelity and relaxation time arises from the trade-off between predominantly cavity induced loss (at a rate $\kappa$) near resonance and intrinsic qubit dissipation ( at a rate $\Gamma$) far off resonance. Interestingly, we observe another peak (dip) in the steady state fidelity (relaxation time) at $\delta=-B_0/2$ corresponding to a two photon resonance. This qualitative behavior remains even outside the adiabatic limit probed in the experiment, as shown in the following section. 

The fast-ramp out measurements, shown in Fig.~\ref{fig:hourglass}, perform a measurement in the bare basis of the qubit: $\langle\sigma_z\rangle$. In this basis, the instantaneous eigenstates of $H_B^+$ acquire a periodic time-dependence as the applied field rotates in the $xz$-plane of the Bloch sphere. Neglecting admixture from the one-photon sector in the cavity and imperfect diabaticity of the ramp out, the value of $\langle\sigma_z\rangle$ obtained from the fast ramp-out measurement is approximately $P(\phi_+)\sin(\omega_{\mathrm{mod}}t)-P(\phi_-)\sin(\omega_{\mathrm{mod}}t)$. Far from resonance, states initially prepared in $\ket{\phi_-}_0$ versus $\ket{\phi_+}_0$ therefore evolve with a $\pi$-phase shift, as shown in Fig.~\ref{fig:hourglass}(a). The oscillatory response damps out as the both states decay to a mixed state. Just as with the slow ramp-out data, the decay time is significantly longer than the native phase-coherence times, indicating a reduced sensitivity to frequency noise on the qubit transition. 

Near resonance, if the qubit is initialized in $\ket{\phi_-}_0$, it will remain close to $\ket{\phi_-}$ indefinitely, leading to persistent oscillations [blue curves in Fig.~\ref{fig:hourglass}(b-c)]. However, qubits initially prepared in $\ket{\phi_+}_0$ behave qualitatively differently. The initial oscillation damps out, followed by a high-contrast persistent oscillation with the same phase as the $\ket{\phi_-}$ state [orange curves in Fig.~\ref{fig:hourglass}(b-c)]. These steady-state oscillations persist for timescales far exceeding both the qubit and photonic decoherence times of the system. The crossover from an out-of-phase oscillations to a persistent in-phase oscillation is most clearly visible at intermediate detuning, such as that shown in Fig.~\ref{fig:hourglass}(b), where $\delta \approx 2g$. 

Due to increased loss from the cavity near resonance during the adiabatic state preparation phase, the states starting from both $\ket{g}$ or $\ket{e}$ end up with significant weight on $\ket{\phi_-}$, even at $t=0$ (yellow curves in Fig.~\ref{fig:adiabatic_detuning_scan}).  
When this weight is near or above $50\%$, the corresponding fast ramp-out data is not expected to display initial out-of-phase oscillations, as seen in Fig.~\ref{fig:hourglass}(c).

\section{Dissipative Stabilization Beyond the Adiabatic Regime}\label{sec:beyondAdiabatic}

\begin{figure}[t]
\centering
		\includegraphics[width=\linewidth]{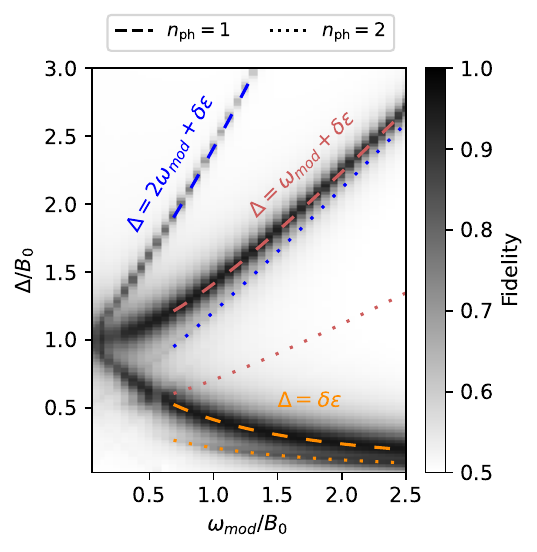}
    \vspace{-0.2cm}
	\caption{\label{fig:nonadiabatictheory} 
    \textbf{Stabilization to a target quasienergy state in various drive regimes.} 
    The steady state in the driven and dissipative spin-cavity system has nearly unit fidelity with the quasienergy state \(\ket{\phi_-}\) [Eq.~\eqref{eqn:fidelity}] (indicated by a dark color) when a resonance condition is met [lines, Eq.~\eqref{Eq:GeneralizedResonanceCond}]. The resonances with $m=0,1,2$, and $n_\mathrm{ph}=1$ (dashed lines) and $n_\mathrm{ph}=2$ (dotted lines) all lead to stabilization, with the $m=0, n_\mathrm{ph}=1$ line being the most robust. 
\emph{Parameters:} $g/B_0 = 0.05$, $\kappa/B_0 = 0.05$ and $\Gamma = \kappa/20$. See Appendix~\ref{appendix:numerics_details}. 
    } 
\end{figure}

Floquet quasienergy states can be stabilized with static dissipation in any frequency regime. 
Away from the adiabatic limit, the quasienergy states are no longer close to instantaneous eigenstates. Nevertheless, the mechanism outlined in Sec.~\ref{sec:model} and in Fig.~\ref{fig:intuition_cartoon}(b) stabilizes a target quasienergy state when the right resonance condition is met. In this section, we formalize the heuristic argument in Sec. \ref{sec:model}, and test its myriad predictions with numerical simulations (Fig.~\ref{fig:nonadiabatictheory}).

In the strong drive and strong coupling regime [Eq.~\eqref{Eq:HierarchyScales}] of the augmented spin-cavity system in Eq.~\eqref{eqn:experimentH}, a target quasienergy state is stabilized if the resonance condition (approximately) holds for integer $m$:
\begin{equation}
     \Delta =  m \omega_{\mathrm{mod}} + \delta \epsilon,
    \label{eqn:quasi_resonance}
\end{equation}
where $\delta \epsilon =  \epsilon^{(0)}_+ - \epsilon^{(0)}_-$ is the quasienergy difference in a Floquet zone. Near resonance, the states of the joint system are not close to separable product states. Instead, to leading order in the spin-cavity coupling $g$, the qubit-cavity system has hybridized quasienergy states
\begin{equation}
\label{Eq:HybridizedQE}
    \ket{\phi_{\text{qb-cav}}(t)} \approx \frac{1}{\sqrt{2}}\left(\ket{\phi^{(0)}_-(t), n+1} \pm \ket{\phi^{(m)}_+(t), n} \right),
\end{equation}
in every photon number sector $n$, except for one unhybridized state, \(\ket{\phi^{(0)}_-(t), 0}\).
Reinstating photon loss, all states in Eq.~\eqref{Eq:HybridizedQE} decay (including crucially the states \(\ket{\phi^{(m)}_+(t), 0}\), which acquire a one-photon character through the hybridization) at a rate set by $\kappa$. If $\kappa$ exceeds the intrinsic qubit decay rate $\Gamma$, then this process stabilizes \(\ket{\phi_-(t), 0}\) with high fidelity on a time scale set by $\kappa^{-1}$.

Figure~\ref{fig:nonadiabatictheory} numerically confirms that a representative quasienergy state $\ket{\phi_-(t)}$ is stabilized whenever the resonance conditions are met. We simulate the augmented spin-cavity system with the Hamiltonian in Eq.~\eqref{eqn:experimentH} and dissipation using a Floquet-Lindblad master equation. The simulation includes two jump operators: $a$ for photon loss, and $\sigma^-$ for qubit relaxation to the $\ket{g}$ state. See Appendix \ref{appendix:numerics_details} for more details. The $x$-axis varies the frequency of the drive on the qubit, going from adiabatic on the left to high frequency on the right, while the $y$-axis varies the effective cavity frequency. The color quantifies the period-averaged trace overlap between the steady state of the qubit-cavity system $\rho_\mathrm{SS}(t)$ and the target quasienergy state:
\begin{align}
    F = \frac{1}{T_{\mathrm{mod}}} \int_0^{T_{\mathrm{mod}}} \mathrm{d}t\, \mathrm{Tr} \left[\rho_\mathrm{SS}(t) (\ket{\phi_-(t)}\bra{\phi_-(t)} \otimes \mathds{1})\right].
    \label{eqn:fidelity}
\end{align}
We immediately see that when the resonance conditions hold for $m=0,1,2$ (dashed lines), the steady state has nearly unit overlap with the target quasienergy states (black color). At generic detunings however, this overlap is close to one-half, as the steady state is close to being maximally mixed between the two quasienergy states of the isolated qubit system. In Appendix~\ref{appendix:numerics_details}, we also demonstrate that the timescale for decay to the steady state is shortest, of order $\kappa^{-1}$, when the resonance conditions hold. 

There are several other features visible in Fig.~\ref{fig:nonadiabatictheory} that can be simply understood within the resonance-based picture of Fig.~\ref{fig:intuition_cartoon}(b). First, the resonance condition need not be perfectly met; the width of the black curves quantifies how close $\Delta$ needs to be the dashed curve. This is because the $-$ and $+$ quasienergy states hybridize between number sectors of the cavity as long as the matrix element connecting the states, $|\tilde{H}^{(m0)}_{+-}|$, exceeds their energy difference. A related second observation is that the larger the value of $m$, the less broad the black curve. That is, the resonance condition has to be met more stringently as $m$ increases. This follows from the dependence of $|\tilde{H}^{(m0)}_{+-}|$ on $m$. The matrix element is given as,
\begin{equation}
    \tilde{H}^{(m0)}_{+-} = \frac{1}{T_{\mathrm{mod}}} \int_0^{T_{\mathrm{mod}}} \mathrm{d}t\, \bra{\phi^{(m)}_+(t), n} V \ket{\phi^{(0)}_-(t), n+1},
    \label{eqn:quasi_mat_ele}
\end{equation}
which is the time-averaged matrix element of the qubit-cavity coupling \(V = g(a^\dagger \sigma^- + a \sigma^+)\) between the quasienergy states in adjacent photon number sector. Taking \(m\) to be large causes the integral in Eq.~\eqref{eqn:quasi_mat_ele} to rapidly oscillate and for $|\tilde{H}^{(m0)}_{+-}|$ to become (exponentially) small in $m$. Further, even when the resonance condition is exactly met, the smallness of $|\tilde{H}^{(m0)}_{+-}|$ takes the system out of the strong coupling regime at large enough $m$. This changes the rate of decay of the state $\ket{\phi_+,0}$ to a rate set by Fermi's golden rule, $|\tilde{H}^{(m0)}_{+-}|^2/\kappa$, and significantly slows down the stabilization process. Indeed, for fixed rate of intrinsic spin dissipation $\Gamma$, the stabilization
would disappear at large enough $m$ as spin dissipation eventually dominates over the stabilizing process. 
For the parameters in Fig.~\ref{fig:nonadiabatictheory}, stabilization disappears for $m\geq 3$ (see Appendix~\ref{appendix:numerics_details} for more details).

A third feature in the figure is that there are faint black lines not given by the resonance condition in Eq.~\eqref{eqn:quasi_resonance} (shown as dotted lines). These are also a consequence of dissipative stabilization, but now by \emph{two-photon} resonances. For example, the dotted line in orange is the condition that $\ket{\phi_+,0}$ and  $\ket{\phi_-,2}$ are resonant. The generalized resonance criterion
\begin{align}
\label{Eq:GeneralizedResonanceCond}
 n_\mathrm{ph} \Delta =  m \omega_{\mathrm{mod}} + \delta\epsilon ,
\end{align}
with $m\in \mathbb{Z}$ and $n_\mathrm{ph} \in \mathbb{Z}^+$ accounts for these faint lines. At first order in $g$, only those resonances with $n_\mathrm{ph}=1$ are accessible, as the Jaynes-Cummings term couples states which differ by exactly one photon. Indeed, naive perturbation theory suggests that resonances between quasienergy states that differ by \(n_\mathrm{ph}\) photons in the cavity come at order \(n_\mathrm{ph}\) in $g$. Analogously to the high-order Floquet resonances indexed by \(m\), these higher photon resonances become narrower and and harder to access with increasing \(n_\mathrm{ph}\). This explains why the numerical simulations find no evidence for stabilization by three photon resonances.  
Nevertheless, dissipative stabilization from the two-photon resonance is quite robust; we observe stabilization by the two-photon resonance in the experiment as well, see Appendix~\ref{appendix:timescales_and_contrast}.

We close this section with a discussion of the adiabatic limit in the theoretical framework of this section. Observe that the dashed and dotted lines indicating the one and two photon resonance conditions are not drawn for values below $\omega_\mathrm{mod}/B_0 \approx 1$, while the black curves smoothly continue and meet near $\Delta/B_0 \approx 1$. Below $\omega_\mathrm{mod}/B_0 \approx 1$, the quasi-energies in a single Floquet zone are separated by $\approx \omega_\mathrm{mod}$, but the matrix element $|\tilde{H}^{(00)}_{+-}|$ is sufficiently small that the system is not in the strong coupling regime and dissipative stabilization disappears. The matrix element $|\tilde{H}^{(m0)}_{+-}|$ is instead maximal between copies of the quasienergy states that are nearly $B_0$ apart in quasienergy, or between Floquet zones separated by $m \approx B_0/\omega_\mathrm{mod}$. This reconciles the resonance criteria in Eq.~\eqref{Eq:AdLimitResonance}  and Eq.~\eqref{Eq:GeneralizedResonanceCond}, and explains why the darkest black curves meet as $\omega_\mathrm{mod}/B_0 \to 0$ at $\Delta/B_0=1$. The fainter secondary curves meet at \(\Delta/B_0 = 1/2\), corresponding to the two-photon resonance condition in the adiabatic limit.

\section{Dissipative Stabilization of a Topological Photon Pump}\label{sec:pumping}

Quantum states can serve as resources for classically difficult tasks, making reliable and stable preparation of these states a useful module in a variety of quantum engineering protocols. The most familiar examples of such resource states are static---a given scheme would call for the preparation of a state \(\ket{\psi_{\mathrm{resource}}}\), which does not evolve except when acted on by the protocol~\cite{Giovannetti2011:metrology,Toth2014:metrology_info,Briegel2009:MBQC,Knill2004:magic,Bravyi2005:magic}.
However, some schemes explicitly call for time-varying periodic states, \(\ket{\psi_{\mathrm{resource}}(t)}\), and utilize this time dependence as part of their dynamical protocol~\cite{Martin:2017aa,Crowley:2019_classification,Nathan:2019_drivendiss,PSAROUDAKI2021,Long:2022boost}. These periodic resource states can be stabilized by our methods.

Cavity state boosting~\cite{Long:2022boost} is a topological pumping protocol which can be used to coherently translate, or \emph{boost}, the quantum state of a cavity in the Fock basis. This enables the preparation of highly excited non-classical states, such as Fock states, if lower photon number states can be reliably prepared (see Refs.~\cite{Varcoe:2000aa,Hofheinz:2008wb, Wang:2008aa, Heeres:2017aa, Rivera:2023aa} for other methods to prepare Fock states). More generally, given an initial cavity state written in terms of Fock states
\begin{equation}
    \ket{\psi_{0}} = \sum_{n} c_n \ket{n},
\end{equation}
cavity boosting produces evolved states
\begin{equation}
    \ket{\psi(MT_{\mathrm{mod}})} \approx \sum_{n} c_n \ket{n + M}
    \label{eqn:boosting}
\end{equation}
at a sequence of predictable times \(t=MT_{\mathrm{mod}}\), where \(T_{\mathrm{mod}}\) is the period of a time-varying resource state and \(M\) is an integer. 

The required resource state is a strongly varying qubit state that covers the Bloch sphere~\cite{Martin:2017aa,Crowley:2019_classification}. When this state is chosen correctly and coupled to the cavity with exchange (Jaynes-Cummings) interactions, the influence of the qubit on the cavity is to effectively implement a Thouless pump in photon number space. On average, every cycle of the qubit state pumps a integer number \(C\) of photons into the cavity, leading to an average increase in the photon number with the quantized rate
\begin{equation}
    \lim_{t\to\infty}\frac{1}{t}\int_0^t \langle\dot{n}(t')\rangle \mathrm{d}t' = \frac{C}{T_{\mathrm{mod}}},
    \label{eqn:pump_rate}
\end{equation}
where \(\dot{n} = \mathrm{d}n/\mathrm{d}t\) is the average photon number current into the cavity~\cite{Nathan:2019_drivendiss}.
Henceforth, we take \(C = 1\).
Further, Ref.~\cite{Long:2022boost} showed that at times \(t^*\) which are close to being a common period of the cavity and the qubit, \(t^* \approx N(2\pi/\Delta_b) \approx M T_{\mathrm{mod}}\), the cavity state rephases into a boosted copy of the initial state [Eq.~\eqref{eqn:boosting}]. The cavity frequency, $\Delta_b$, is defined in Eq.~\eqref{Eq:HBoost}.

Reference~\cite{Long:2022boost} appealed to an adiabatic control scheme for the resource qubit. By implementing a strong, slow drive on the qubit, the back-action of the cavity on the qubit is negligible except when they need to exchange energy. This necessitates an operating regime where the pumping rate---given by \(1/T_{\mathrm{mod}}\)---is very low, which in turn limits the size of a cavity state that can be prepared in a time- or loss-limited experiment. The introduction of an auxiliary lossy element can stabilize the required resource state, lending additional robustness to the boosting protocol and hence allowing for faster, more reliable preparation of excited cavity states.

More concretely, stabilized boosting can be achieved in a model with two cavities coupled to a periodically driven qubit,
\begin{multline}
    H_{\mathrm{boost}}(t) = \frac{1}{2}\vec{\sigma}\cdot\vec{B}_{\mathrm{boost}}(t) + \Delta_{\mathrm{b}} a_{\mathrm{b}}^\dagger a_{\mathrm{b}} + \Delta_{\mathrm{s}} a_{\mathrm{s}}^\dagger a_{\mathrm{s}} \\
    + g_{\mathrm{b}}(a_{\mathrm{b}}^\dagger \sigma^- + a_{\mathrm{b}} \sigma^+)
    + g_{\mathrm{s}}(a_{\mathrm{s}}^\dagger \sigma^- + a_{\mathrm{s}} \sigma^+).
    \label{Eq:HBoost}
\end{multline}
The boost cavity---with frequency \(\Delta_{\mathrm{b}}\), annihilation operator \(a_{\mathrm{b}}\), and coupling \(g_{\mathrm{b}}\) to the qubit---holds the non-classical state which is to be boosted, and ideally would be lossless. The stabilizing cavity---with label \(\mathrm{s}\) to differentiate it from the boost cavity---stabilizes the resource qubit. It experiences photon loss with a power decay rate \(\kappa_{\mathrm{s}}\) (which we model through Lindblad evolution with a jump operator \(a_{\mathrm{s}}\)). 

\begin{figure*}
\centering
		\includegraphics[width=\textwidth]{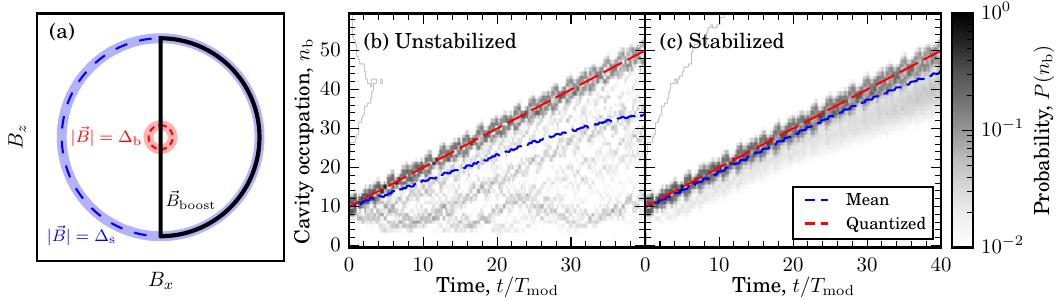}
	\vspace{-0.6cm}
	\caption{\label{fig:boosting} 
    \textbf{Stabilized cavity state boosting.} 
    (a)~The \(\vec{B}\) field for the boosting protocol traces out a semicircle of radius \(B_0\) (black) on the $B_x - B_z$ plane with period \(T_{\mathrm{mod}}\). On the blue dashed circle, the auxiliary cavity s is resonant with the qubit, while on the red dashed circle, the boost cavity b is resonant with the qubit. (b, c) Plots of the Fock state occupation of the boost cavity (\(P(n_{\mathrm{b}}) = \bra{n_{\mathrm{b}}} \rho_{\mathrm{b}}(t) \ket{n_{\mathrm{b}}}\) where \(\rho_{\mathrm{b}}(t)\) is the reduced density matrix for the boost cavity) without (b) and with (c) the auxiliary cavity as a function of protocol duration.
In (b), nonadiabatic excitations of the qubit produce significant population in the cavity at small photon number, indicated by a dark color in the figure. As a result, the expected photon number in the cavity state (blue dashed) is significantly below the ideal, quantized, value [red dashed, Eq.~\eqref{eqn:pump_rate}]. (c)~With the auxiliary cavity, nonadiabatic excitations of the qubit are suppressed, resulting in $\langle n_b \rangle$ becoming closer to the quantized value and cavity states that are closer to the desired high photon number Fock state at the special rephasing times $t^*$.
\emph{Parameters:} \(B_0/g_{\mathrm{b}} = 20\), \(\omega_{\mathrm{mod}}/g_{\mathrm{b}} = 1.5\), \(\Delta_{\mathrm{b}}/\omega_{\mathrm{mod}} = (1+\sqrt{5})/2\), \(\Delta_{\mathrm{s}}/B_0 = 1\), initial state \(\ket{g, n_{\mathrm{b}} = 10, n_{\mathrm{s}}=0}\); (b)~\(g_{\mathrm{s}}/g_{\mathrm{b}} = 0\); (c)~\(g_{\mathrm{s}}/g_{\mathrm{b}} = 1\), photon loss from the lossy cavity with rate \(\kappa_{\mathrm{s}}/g_{\mathrm{b}} = 1\) is modeled with a jump operator \(a_{\mathrm{s}}\) in a Lindblad equation, other dissipative processes are neglected. }
\end{figure*}

We choose a qubit drive 
\begin{equation}
    \vec{\sigma}\cdot\vec{B}_{\mathrm{boost}}(t) = B_0 \left[ \max\{0, \sin(\omega_{\mathrm{mod}} t)\} \sigma_x - \cos(\omega_{\mathrm{mod}} t) \sigma_z \right].
    \label{Eq:DrivePart}
\end{equation}
The vector \(\vec{B}_{\mathrm{boost}}(t)\) traces out a semicircle of radius \(B_0\) in the \(B_x-B_z\) plane [Fig.~\ref{fig:boosting}(a)].

Cavity state boosting (and topological photon pumping more broadly) requires the boost cavity energy and drive frequencies to be much smaller than the qubit energy and interaction: \(\Delta_{\mathrm{b}}, \omega_{\mathrm{mod}} \ll B_0, g_{\mathrm{b}}\sqrt{n_{\mathrm{b}}}\)~\cite{Nathan:2019_drivendiss}.
Meanwhile, stabilization requires a resonance condition between the qubit quasienergy (which can be replaced by the instantaneous energy near the adiabatic limit) and the stabilizing cavity: \(|\Delta_{\mathrm{s}}-B_0|\lesssim g_{\mathrm{s}}\).

In order for the boost cavity to not couple strongly to the stabilizing cavity, and inherit its photon loss, we also demand that they are far detuned: \(|\Delta_{\mathrm{s}} - \Delta_{\mathrm{b}}| \gg g_{\mathrm{b}}, g_{\mathrm{s}}\).

Provided these hierarchies of scales are satisfied, there is a simple toy picture for boosting. At the beginning of the protocol (\(t=0\)) the cavity is initialized in an accessible Fock state \(\ket{n_0}\) and the qubit is prepared in its ground state \(\ket{g}\), which is an instantaneous eigenstate of Eq.~\eqref{Eq:DrivePart}. The qubit then adiabatically follows the rotating external field \(\vec{B}_{\mathrm{boost}}(t)\) from the negative \(B_z\) direction to the positive direction, and ends in the \(\ket{e}\) state. As there is a large \(B_x\) field on the qubit when \(B_z\) crosses \(\Delta_b\), the qubit and cavity remain far detuned and unentangled, though the boost cavity state experiences a mean displacement field from the qubit \(g_b(a_b^\dagger \langle \sigma^-\rangle + \mathrm{H.c.})\). Then, in the vertical segment of the semicircle protocol, \(\vec{B}_{\mathrm{boost}}\) is slowly ramped through an avoided level crossing with the boost cavity, causing an adiabatic transition from the \(\ket{e,n}\) state to the state \(\ket{g,n+1}\) as the cavity absorbs a qubit excitation. The protocol is then repeated at a fixed rate $1/T_\mathrm{mod}$. While the displacement field causes the cavity to deviate from a Fock state at most stroboscopic times, at the special rephasing times \(MT_{\mathrm{mod}}\) the total displacement averages to zero and an approximate Fock state is recovered.

The introduction of the auxiliary stabilizing cavity suppresses nonadiabatic excitation of the qubit along the curved part of the semicircle protocol through precisely the same mechanism as Sec.~\ref{sec:expt}. Without this, the qubit has a higher probability to flip to the \(\ket{g}\) state by the beginning of the vertical segment, and then absorb photons from the cavity by adiabatically following the opposite branch of the avoided crossing, transitioning from \(\ket{g,n}\) to \(\ket{e,n-1}\). This nonadiabatic process results in a cavity state which: (i)~has a lower mean photon number than predicted by the ideal, quantized, rate of energy pumping in Eq.~\eqref{eqn:pump_rate}; and (ii)~is no longer a perfect Fock state, instead having some nonzero variance in its photon number distribution.

We numerically verify the quantitative improvement of boosting due to introducing coupling to the auxiliary lossy cavity in Figs.~\ref{fig:boosting}(b,c) using a Lindblad master equation simulation. The stabilized boosting protocol produces (i)~an average pumping rate which is closer to the quantized expected value of one photon per period [Eq.~\eqref{eqn:pump_rate}]; and (ii)~a boosted Fock state with a smaller photon number variance, despite the introduction of dissipation. This is shown quantitatively in Appendix~\ref{appendix:numerics_details}, where we also discuss details of our numerical simulations.


\section{Conclusion and Outlook}
\label{sec:conclusion}

We have demonstrated a simple and fully autonomous scheme to stabilize target states in Floquet systems. To our knowledge, this is the first such scheme which applies \emph{both} with strong driving, for which laboratory-frame ground and excited states are strongly mixed during a Floquet cycle, \emph{and} across all drive frequency regimes, from the low-frequency adiabatic limit all the way up to high drive frequencies exceeding all other energy scales in the problem. Our method relies on the natural static coupling between the Floquet system of interest and an un-modulated auxiliary system, and on a separation of energy scales which allows a target quasienergy state $\ket{\phi_-(t)}$ to hybridize weakly with the auxiliary compared to all other quasienergy states. Dissipation in the auxiliary then relaxes the combined system toward a steady state with $\ket{\phi_-}$ in the Floquet system and vacuum in the auxiliary.

We have provided combined experimental and theoretical evidence for stabilization in a strongly-driven two-level system coupled to a lossy harmonic oscillator. Experimentally, we study a driven flux-tunable transmon qubit coupled to fixed-frequency microwave resonator, and observe robust steady state stabilization of the $\ket{\phi_-(t)}$ state in the adiabatic regime. 
Our numerical simulations demonstrate stabilization in all frequency regimes. In addition to this proof-of-principle two-level system example, we numerically studied a two-cavity model in which this stabilization effect improves the performance of a topological photon pump.

The combined experimental and theoretical results demonstrate that the stabilization scheme effectively combats heating processes in Floquet systems which arise due to admixture of excited states with finite decay and dephasing rates.
On the other hand, the Hilbert space of a driven two-level system model is too small to exhibit non-trivial heating due to leakage out of the desired operational subspace caused by higher-order drive processes or due to interaction effects with more degrees of freedom, so the results on the phenomenology of $H_B^+$ do not directly imply robustness against these kinds of heating. 

However, there are two aspects of our results that suggest that the same scheme can combat these other kinds of heating. First, our experimental implementation is carried out on a transmon qubit, which is not a true two-level system. In fact, the transmon is \emph{notorious} for being only loosely describable as a two-level system due to its relatively low anharmonicity of $100-250$~MHz (or roughly $2\%-5\%$). The data in Fig.~\ref{fig:hourglass} shows persistent oscillations for $100\ \mu$s and over $100$ drive periods without signs of uncompensated heating due to the higher levels of the transmon, despite applied drive strengths which are roughly one third of the anharmonicity. Second, the stabilized topological photon pump described in Sec.~\ref{sec:pumping} uses a second lossy cavity to stabilize a Floquet system consisting of \emph{both} a qubit and a cavity. 
The presence of autonomous stabilization in both of these cases despite the larger Hilbert space is a strong indicator that it is possible to suppress Floquet heating in general using our simple scheme, rather than the contributions of excited-state lifetime alone. 
However, a full characterization of the capability of autonomous stabilization with a static auxiliary remains an outstanding question.

A major advantage of dissipative stabilization methods, such as those we present here, is that they are robust to slight imperfections and variations of parameters.
While \emph{optimal} stabilization---with the fastest possible stabilization time and highest possible steady-state purity---relies on exact resonances in the augmented spin-cavity system, significant stabilization can be achieved for a much wider range of parameters. 
This includes both varying parameters within our studied driving protocol, but also altering the protocol completely, or changing experimental systems entirely.
Indeed, supplementary experimental data (Appendix~\ref{appendix:timescales_and_contrast}) shows that significant stabilization is observed for \emph{all} drive parameters in the range $|\delta| \lesssim g$. As the transmon decay rate in our device is only a fourth of the photon decay rate, using an even lossier cavity (up to $\kappa \approx g$) would result in stabilization for an even wider range of detunings. Varying our drive protocol, we also find that elliptical drive fields, which are only instantaneously near-resonant during part of the drive cycle, also exhibit stabilization.

The robustness of dissipative stabilization opens the door to potential applications in settings far beyond the experiment performed here.
For instance, a single sufficiently lossy auxiliary degree of freedom could be used to counteract heating into \emph{all} excited states within a range of roughly $g$ of the resonance, and hence systematically remove population from targeted Floquet bands.
If either the Floquet band gap or the cavity frequency can be varied in time, then the entire Floquet band could be emptied of excitations. Using similar time-dependent schemes, entangled states can be prepared, simplifying current schemes that target multiple resonances at the same time~\cite{Shankar:2013, Brown:2022}.

Alternatively, an auxiliary consisting of a band-pass filter, rather than a single cavity, could be used to enhance loss for a large number of Floquet states~\cite{Wang:2024_0pi}.
Implementations along these lines would give the ability to, for example, cool directly into the ground state of a Floquet engineered Hamiltonian~\cite{Seetharam:2015aa}, rather than relying on adiabatic state preparation, which is slow and fails near phase transitions.
We note that our natively-Floquet dissipative scheme is well-adapted to emerging qubit designs which utilize periodic driving explicitly~\cite{Wang:2024_0pi}.

When applied to a many-body periodically driven system in the high-frequency regime, the scheme discussed in this article can cool the system to close to the ground state of a gapped effective Hamiltonian~\cite{Petiziol:2022aa, Schaefer:2024}.
The effective Hamiltonian controls stroboscopic dynamics and can be derived in a high-frequency expansion~\cite{Bukov2015}.
In the absence of the dissipative element, the time evolution of the system is captured by a Gibbs state with a slowly increasing temperature as a function of time. The dissipative element removed entropy from this Gibbs state and combats the slow heating for optimal choices of $g, \kappa$. When the effective Hamiltonian is gapped, our upcoming work~\cite{Schaefer:2024} shows that low temperatures smaller than the gap of the effective Hamiltonian can be reached in the steady state. 



\par
\begin{acknowledgments}
We thank Maya Amouzegar for assistance with device fabrication, Benjamin Cochran for contributions to device modeling, IBK Adisa for assistance with early measurements; and Vedika Khemani, Chris Laumann, and Paul Schindler for helpful comments. 

During completion of this manuscript, we became aware of an unrelated proposal~\cite{mikeK} in which the ability to achieve resonance in a Jaynes-Cummings model using strong Rabi drives was harnessed for a topological pump.

This work was supported by ARL (Grants no. W911NF-19-2-0181 and W911NF-17-S-0003), the University of Maryland, and AFOSR (Grants No. FA9550-20-1-0235 and FA9550-24-1-0121). MR received support from the National Science Foundation (PFC at JQI Grant No.  PHY-1430094), the Laboratory for Physicsal Sciences graduate fellowship, and ARCS. QY received support from AFOSR Grant No. FA9550-21-1-0129. DL was supported by the Laboratory for Physical Sciences, a Stanford Q-FARM Bloch Postdoctoral Fellowship, and the Packard Foundation through a Packard Fellowship in Science and Engineering (PI: Vedika Khemani). AC thanks the Max Planck Institute for the Physics of Complex Systems for its hospitality

\end{acknowledgments}


\bibliographystyle{apsrev4-1-prx}
\bibliography{refs.bib}

\newpage
\onecolumngrid
\appendix

\section{Device Fabrication and Design}\label{appendix:fabrication}

The device used for the experiment consists of a tunable transmon coupled to two coplanar waveguide (CPW) cavities which we refer to as the readout and main cavities respectively. Table \ref{tab:qubit_params} shows all the extracted experimental parameters. An on-chip flux-bias line (FBL) along with an external magnet provide AC and DC control of the qubit frequency by changing the flux through the DC SQUID forming the inductive portion of the transmon. Coherent control and readout are performed through the readout cavity, which is far detuned from the operating qubit frequency. The main cavity, much closer in frequency to the qubit operating point, introduces the external loss through hybridization with the qubit. 

The device was fabricated on a sapphire substrate with a tantalum (Ta) superconducting ground plane. The qubit paddles and resonators were defined using standard photolithography techniques followed by a wet etch. Finally, the qubit Josephson junctions were fabricated using electron-beam lithography followed by double angle aluminum evaporation and oxidation. 

\begin{table}[h]
    \centering
    \begin{tabular}{|c|c|c|}
    \hline
    Parameter & Symbol & Value \\
       \hline \hline
       Qubit g-e frequency & $\omega_q$/(2$\pi$)  & 3.9-7.4 GHz\\
       \hline
       Qubit anharmonicity & $\alpha$/(2$\pi$)    & 240 MHz \\
       \hline
       Qubit-main cavity coupling & $g_m$/(2$\pi$)       & 13 MHz\\
       \hline
       Qubit-readout cavity coupling & $g_r$/(2$\pi$)       & 90 MHz\\
       \hline 
       Qubit decay rate & $\Gamma_q$/(2$\pi$) & 13.8 kHz \\
       \hline \hline
       Readout cavity frequency & $\omega_r$/(2$\pi$)  & 7.492 GHz\\
       \hline
       Readout cavity linewidth & $\kappa_r$/(2$\pi$)  & 350 kHz\\
       \hline \hline 
       Main cavity frequency & $\omega_m$/(2$\pi$)  & 5.04 GHz\\
       \hline
       Main cavity linewidth & $\kappa_m$/(2$\pi$)  & 84 kHz\\
       \hline
    \end{tabular}
    \caption{Device Parameters. The qubit and main cavity parameters are acquired with the qubit at 4.9 GHz to isolate the individual decay rates and linewidths.
    $\Gamma_q$ is defined as $1/T_1$, the population decay rate of the qubit from the excited state without Purcell loss into the main cavity. 
  }
    \label{tab:qubit_params}
\end{table}

\section{Synthesizing a Rotating Field}\label{appendix:B_field_synthesis}

A rotating magnetic field, or a circularly polarized magnetic field, consists of two oscillating components with a $\pi$/2 phase shift between them. We synthesize such an effective rotating field in the $xz$ plane with amplitude $B_0$ and rotation rate $\omega_{\mathrm{mod}}$,
\begin{equation}\label{eqn:Bwish}
    \vec{B}(t) = B_0 \begin{pmatrix} \cos(\omega_{\mathrm{mod}} t) \\ 0 \\ \sin(\omega_{\mathrm{mod}} t) \end{pmatrix},
\end{equation}
using a microwave drive ($X$-drive) to provide the $x$ component and frequency modulation of the qubit ($Z$-drive) to provide the $z$ component of the field. Here we sketch the general theoretical framework for producing an effective magnetic field of this form for a transmon qubit, which does not natively exhibit a magnetic moment. Technical details of the required control signals and calibration protocols are described in \cite{syntheticFields}.

The transition frequency of a transmon formed using a DC SQUID in parallel with a capacitor is controlled by the external flux threaded through the SQUID \cite{Blais_2021, Krantz:2019}. The $z$-component of the desired drive is therefore relatively straightforward to implement. A DC bias, set by an external magnet to reduce noise, is used to set the desired average qubit frequency, and an AC bias is applied using an on-chip flux-bias line to modulate the qubit frequency about this value:
\begin{equation}
    H = \frac{1}{2}\sigma_z \left(\omega_{q_0}+B_0 \sin(\omega_{\mathrm{mod}}t)\right),
\end{equation}
where $\omega_{q_0}$ is the qubit frequency set by the DC bias, and $B_0$, $\omega_{\mathrm{mod}}$ are the amplitude and rotation frequency of the applied magnetic field respectively. 
The DC bias point is maintained by an active servo \cite{syntheticFields} to keep the average qubit frequency at the desired point even in the presence of environmental drift. The AC bias line is configured with a $5$~MHz low-pass filter at the base-plate of the dilution refrigerator to cut out high-frequency noise. The required AC current amplitude is calibrated by performing qubit spectroscopy after applying a flux bias pulse. The group delay induced by the filter and line is compensated by adding a calibrated buffer time that matches the time between sending flux bias pulse and observing a response in the qubit spectroscopy. See Ref. \cite{syntheticFields} for details.

In contrast, the $x$-component of the desired effective magnetic field is not realized by applying a laboratory-frame magnetic field.
Instead, we make use of the well-known fact that driving a two-level system  with an oscillatory electric field leads to Rabi oscillations between the $\ket{g}$ and $\ket{e}$ states, which can be mapped to the dynamics of spin in a magnetic field \cite{wallsMilburn}.
The bare laboratory-frame Hamiltonian takes the from
\begin{equation}
    H_{\mathrm{bare}} = \frac{\omega_q}{2}\sigma_z + \vec{E_0}\cdot\vec{d} (a^{\dagger} e^{i\omega_d t}+a e^{-i \omega_d t})(\sigma^- + \sigma^+),
\end{equation}
where $\omega_q, \omega_d$ are the qubit and drive frequencies respectively; $a,\sigma$ are the usual photon and qubit operators; and $\vec{d} $, $E_0$ are the qubit transition dipole moment and the electric field amplitude respectively. Assuming a large coherent-state of the electric field, moving into a co-rotating frame with the applied drive transforms the bare electric Hamiltonian into an effective magnetic Hamiltonian:
\begin{equation}
    H_{\mathrm{eff}} = \frac{\Delta_d}{2}\sigma_z + \frac{\Omega}{2}\sigma_x,
\end{equation}
where $\Delta_d$ = $\omega_q-\omega_d$ is the detuning between the qubit and the drive, and $\Omega$ is the Rabi rate \cite{wallsMilburn}. 

To produce the desired effective magnetic field in Eq.~\eqref{eqn:Bwish}, the detuning with respect to the drive $\Delta_d$ and the Rabi rate $\Omega$ must both be modulated in quadrature. If $\Delta_d(t) = B_0 \sin(\omega_{\mathrm{mod}}t)$ and $\Omega(t) = B_0\cos(\omega_{\mathrm{mod}}t)$, $H_{\mathrm{eff}}$ becomes
\begin{equation}
    H_{\mathrm{eff}} = \frac{\Delta_d(t)}{2}\sigma_z + \frac{\Omega(t)}{2}\sigma_x = \frac{B_0}{2} \left[  \sin(\omega_{\mathrm{mod}}t)\sigma_z + \cos(\omega_{\mathrm{mod}}t)\sigma_x \right]  = \vec{B}(t) \cdot \vec{\sigma},
\end{equation}
as desired. 

In this configuration, $H_{\mathrm{eff}}$ has two instantaneous eigenstates with energies $\pm B_0/2$. The composition of these two eigenstates depends on the quantization axis set by the combination of the two drives. On resonance, only the $\sigma_x$ term remains, and the two eigenstates are equal superpositions of the qubit ground and excited states. Conversely, when the microwave drive is turned off, only the $\sigma_z$ component remains, and the eigenstates of $H_{\mathrm{eff}}$ coincide with the bare qubit eigenstates. As the applied effective magnetic field rotates, the eigenstates interpolate smoothly between these regimes, in exactly the same way as the eigenstates of a true spin-$1/2$ in a true rotating magnetic field. 

Note that at the level of $H_{\mathrm{eff}}$, changing the drive frequency or changing the qubit frequency are physically equivalent. However, this ambiguity is lifted if the qubit-detuning relative to some other component, such as the boosting cavity in Sec.~\ref{sec:pumping}, also has to be modulated. While both methods apply a rotating field in the Bloch sphere of the the qubit, only modifying the qubit frequency will actually translate to modifying its detuning relative to other Hamiltonian elements. 

We highlight this difference by explicitly constructing the rotating-frame Hamiltonian obtained for a qubit coupled to a cavity. The general laboratory-frame Hamiltonian which incorporates both frequency-modulation of the qubit and the drive is:
\begin{equation}
    H = \omega_m a^{\dagger} a + \omega_q(t) \frac{\sigma_z}{2} + E_0(t) \cos[\phi_d(t)](\sigma^-+\sigma^+) + g(a^{\dagger}\sigma^-+a\sigma^+),
\end{equation}
where we define the instantaneous drive frequency as \(\dot{\phi}_d(t) = \omega_d(t)\). The rapid oscillation on the drive term is removed through a rotating frame transformation by \(U = \exp[i\phi_d(t)(a^\dagger a + \sigma_z/2)]\), so that the Hamiltonian becomes (dropping remaining rapidly oscillating terms)
\begin{equation}
    H_{\mathrm{eff}} = \Delta_{dc}(t) a^{\dagger} a + \frac{\Delta_{dq}(t)}{2}\sigma_z+\frac{\Omega(t)}{2}\sigma_x +g(a^{\dagger}\sigma^-+a\sigma^+)\textcolor{red}{,}
\end{equation}
where $\Delta_{dc} = \omega_m-\omega_d(t)$ is the cavity-drive detuning, $\Delta_{dq}(t)=\omega_q(t)-\omega_d(t)$ is the qubit-drive detuning, and \(\Omega(t) = E_0(t)\). While \(\Delta_{dq}(t)\) can be varied by holding the qubit frequency fixed and varying the drive frequency, we observe that the cavity-qubit detuning in this frame, \(\Delta_{dc}(t) - \Delta_{dq}(t) = \omega_m - \omega_q(t)\), can only be varied in time if the physical qubit frequency \(\omega_q(t)\) is modulated.

In addition, modulating the drive frequency instead of the qubit frequency incurs a set of technical difficulties such as frequency dependent attenuation due to in-line amplifiers and the microwave environment, as well as directly populating any other cavities present on the chip if the drive crosses resonance during the protocol (for example the main cavity used for dissipation). We choose to keep the drive frequency $\omega_d$ fixed and apply direct frequency modulation on the qubit using the on-chip flux bias line to circumvent these technical issues. 

A second important hardware constraint to note is that the \emph{sign} of the microwave drive must be reversed during the cycle in order to produce negative values of $B_x$. As a result, a conventional amplitude modulator is not sufficient to realize the required field. We make use of a vector microwave generator with an IQ modulator in order to have control over both the amplitude and the sign of the microwave drive \cite{syntheticFields, Krantz:2019}. We calibrate our $X$-field using standard Rabi oscillation measurements by varying the amplitude of the applied microwave drive and fitting the resulting oscillation frequencies.

\section{Protocol Schedule}\label{appendix:schedule}
\begin{figure*}
    \centering
    \includegraphics[width=\linewidth]{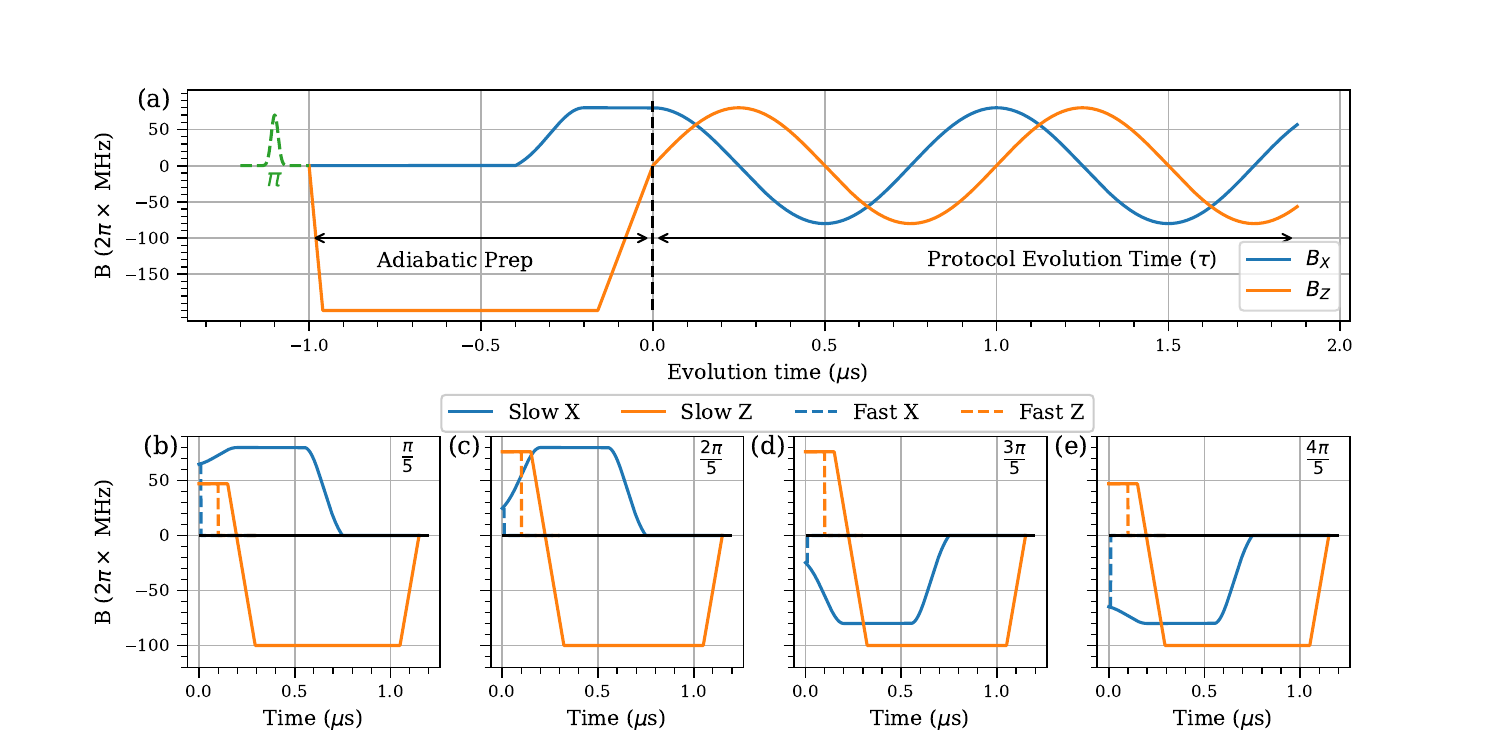}
    \caption{\textbf{Driving Protocol Schedule.}
    (a) State preparation and Hamiltonian evolution sections of the protocol. An optional $\pi$ pulse followed by a ramp to a large $Z$-field (orange trace) adiabatically prepares the qubit in one of the two eigenstates of a large $Z$-field. Increasing the strength of the $X$-field (blue trace) before bringing the $Z$-field to zero allows the qubit state to be adiabatically prepared along the equator of the Bloch sphere. To map out the effect of the rotating magnetic field, we evolve under the driving Hamiltonian for a variable amount of time $\tau$ before ramping out for measurement.
    Sample $B(t)$ for ramping out at different points in the drive cycle are shown in (b)-(e). 
    In the slow ramp out protocol (solid lines), the $X$-field is ramped to its maximum value before the $Z$-field is ramped down to a fixed reference point. This particular order is chosen to minimize spurious diabatic transitions induced by the cavity as the qubit goes through resonance. This scheme adiabatically rotates the qubit and maps the population in the instantaneous basis at time $\tau$ to the population in the $z$ basis. The fast ramp out (dashed lines) consists of a fast shutoff of the $X$-field while holding the $Z$-field constant before shutting off the $Z$-field. The buffer between $X$ and $Z$ shutoffs mitigates the effect of distortions and delays from the $5$~MHz low pass filter on the flux-bias line. This filter also reduces the maximum rate of change of the $Z$-field leading to only an approximate diabatic/instantaneous measurement of the population in the $z$ basis at $t = \tau$. 
    }
    \label{fig:protocol_fig}
\end{figure*}

In this section, we provide more details on the qubit protocol, as well as the two measurement schemes (slow and fast ramp out). The experimental protocols consist of three distinct sections: a state preparation portion, a period of time evolution $\tau$ under the rotating effective magnetic field, and a ramp out/measurement portion. Prior to each experimental run, we choose a desired detuning $\Delta$ that sets the Hamiltonian we want to evolve. We use the external magnet DC bias to set the average qubit frequency $\omega_{q_0}$ to equal $\omega_{d} = \omega_m - \Delta$, and operate the microwave $X$-drive at the same frequency $\omega_{d}$ throughout the protocol. 

The goal of the state preparation portion of the protocol, shown in Fig. \ref{fig:protocol_fig}(a), is to adiabatically prepare the qubit on the equator of the Bloch sphere, either aligned or anti-aligned with the applied magnetic field. Due to the large fields used during the protocol, careful care has to be taken during the initial field ramp to reduce Larmor precession of the qubit state about the driving field. Effectively, we seek to minimize the angle between the effective magnetic field and the qubit state in the Bloch sphere as the amplitude of nutation is set by $|\langle\vec{\sigma}\rangle \times \vec{B}|$.
The qubit is initialized in $\ket{g}$, or in $\ket{e}$ with a $\pi$-pulse, before detuning the qubit by $2\pi\times200$~MHz from the drive frequency by ramping the flux. This creates a large effective magnetic field in the $z$ direction and places the qubit transition frequency far from the cavity so that the qubit is polarized either aligned or anti-aligned to the effective magnetic field ($\ket{\phi_+}$ or $\ket{\phi_-}$). By detuning the qubit from the drive frequency, it is possible to turn on the microwave drive to its nominal value ($B_0 = 2\pi\times80$~MHz) without causing spurious Rabi oscillations. The qubit is ramped back to resonance with a 150~ns ramp, while keeping it in the adiabatic eigenstate of the effective magnetic field it was prepared in. The resulting preparation leads to the qubit pointing along the equator of the Bloch sphere with a pure $X$-field applied via the microwave drive. 

The magnetic field, $B(t)$, is then rotated in the $xz$ plane of the Bloch sphere for a variable time $\tau$ at a rate of $\omega_{\mathrm{mod}}$. The qubit will stay in an instantaneous eigenstate as long as $\omega_{\mathrm{mod}}\ll B_0$. For this work we use an effective magnetic field strengths $B_0$ of $2\pi\times80$~MHz,
and $\omega_{\mathrm{mod}} = 2\pi\times0.75$~MHz, so this condition is well satisfied.

After the desired evolution time, we use two different ramp-out and measurement protocols, shown in Fig. \ref{fig:protocol_fig}(b-e): one, which we call the slow ramp out protocol, designed to adiabatically map the instantaneous eigenstates of the magnetic Hamiltonian back to the bare $\ket{g}$, $\ket{e}$ basis; and a second, called the fast ramp out protocol, in which the effective magnetic field is switched off as quickly as possible and a measurement of $\ket{g}$, $\ket{e}$ basis population is performed. 

The fast ramp out protocol is designed to measure the instantaneous overlap with the $\ket{g}$, $\ket{e}$ basis. In the measurement, even if it is in an instantaneous eigenstate, the qubit state will oscillate between $\ket{g}$ and $\ket{e}$ at a rate set by the $B$-field rotation frequency ($\omega_{\mathrm{mod}}$). The ability of this fast protocol to mimic a perfect diabatic cutoff is limited by the finite bandwidth of the low pass filter on the flux bias line. We mitigate this effect by holding the $Z$-field fixed for $100$~ns while shutting off the $X$-field, which has much higher bandwidth, as quickly as possible. We then ramp the $Z$-field as quickly as possible to zero. The addition of the small intermediate hold time guarantees that the $X$-field is fully off when the slower $Z$-field passes through zero and prevents accidental Rabi flopping during the ramp out. In the absence of any other external perturbations, the hold will cause precession of the qubit state about the $z$ axis, which only modifies phases and does not affect the final $\ket{g}$/$\ket{e}$ population readout. However, near cavity resonance, the qubit-cavity coupling can be non-negligible, and the ramp speed of the $Z$-field attainable with the $5$~MHz filter on the flux bias line is not sufficient to be purely diabatic with respect to the qubit-cavity coupling, $g \approx 2\pi\times13$~MHz. This can lead to coherent excitation exchanges between the qubit and the cavity for measurements in which the qubit crosses resonance with the cavity during ramp out, leading to an only approximate final mapping of the original $\ket{g}/\ket{e}$ populations to the readout result.

The slow ramp out protocol is designed to measure the population in the instantaneous eigenstate basis of the driving Hamiltonian. At the end of the evolution time, the magnetic field is rotated back to pointing down on the Bloch sphere before performing qubit state readout. During the ramp-out phase, the qubit may pass through resonance with both the cavity and the applied microwave drive. To minimize diabatic transitions during the ramp-out phase, the $X$-field is ramped to its maximum value ($\pm$ 2$\pi\times$80 MHz) before the $Z$-field is reduced to maintain a large energy splitting between the instantaneous eigenstates. The qubit is ramped far off resonance to \(-100\)~MHz detuning from the drive frequency, after which the $X$-field is turned off before ramping the qubit back to its original frequency, $\omega_{q_0}$, where readout is performed. This effectively rotates the instantaneous adiabatic eigenstates $\ket{\phi_+}$ and $\ket{\phi_-}$ back to the bare qubit $\ket{g}$ and $\ket{e}$ states for measurement. 

For both ramp out protocols, we average $10^4$ shots to reduce the readout noise from our electronics. The details are discussed in Appendix \ref{appendix:error_bars}. 

\section{Numerical simulations of the augmented spin-cavity and boosting setups}\label{appendix:numerics_details}
\begin{figure}[tb]
    \centering
    \includegraphics[width=0.85\linewidth]{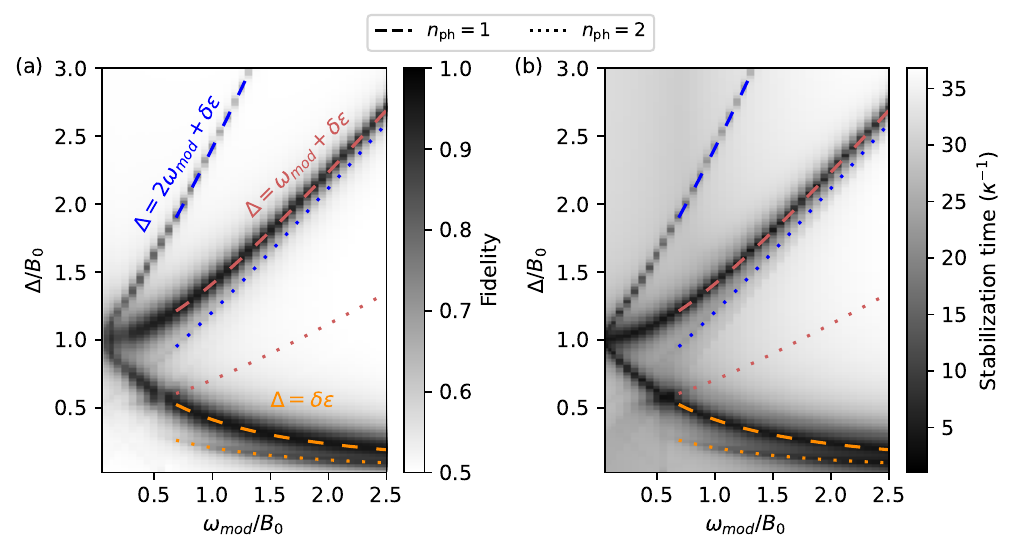}
    \caption{\textbf{Properties of the steady state of the Lindblad-Floquet master equation.} 
    (a)~Reproduced from Fig.~\ref{fig:nonadiabatictheory}. (b)~The time to stabilize into the steady state in units of $\kappa^{-1}$ (color) at different values of the drive frequency (\(x\)-axis) and cavity detunings (\(y\)-axis). The high fidelity between the steady state and \(\ket{\phi_-}\) in (a) is mirrored in (b) by a short stabilization time, on the order of a few $\kappa^{-1}$, of all states to \(\ket{\phi_-}\) in (b). When the resonance condition $n_\mathrm{ph} \Delta  = m \omega_{\mathrm{mod}} + \delta \epsilon$ is met, the state \(\ket{\phi^{(m)}_+,0}\) hybridizes with the state \(\ket{\phi^{(0)}_-,n_\mathrm{ph}}\) and thus decays with a rate set by \(\kappa\), the cavity linewidth. Far from resonance, the time to reach the steady state is the much longer \(1/\Gamma\). \emph{Parameters:} $g/B_0 = 0.05$, $\kappa/B_0 = 0.05$ and $\Gamma = \kappa/20$.} 
\label{fig:nonadiabatictheory_fidelityandlockingtime}
\end{figure}

In this appendix, we describe details of the numerical simulations of driven dissipative dynamics that produced Figs.~\ref{fig:nonadiabatictheory} and \ref{fig:boosting} appearing in the main text. We also present additional data on stabilization rates, the inability of high order resonances to stabilize Floquet states, and the improvement of the topological photon pump when coupled to the auxiliary cavity.

\begin{figure}[tb]
    \centering
    \includegraphics[width=\linewidth]{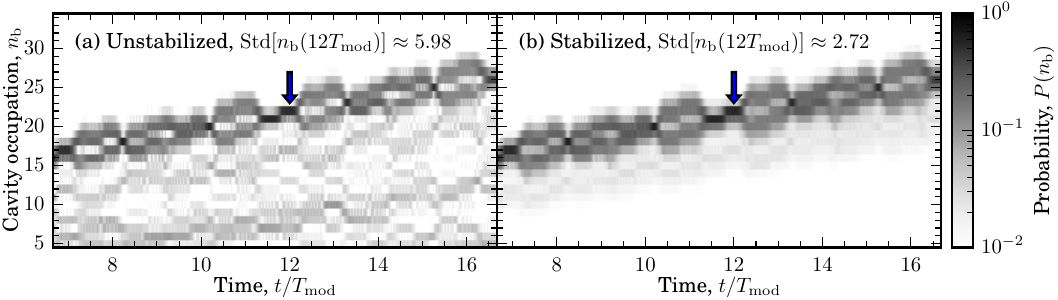}
    \caption{\textbf{Stabilized cavity state boosting.} Plotting the same data as Figs. \ref{fig:boosting}(b-c) on a zoomed-in scale shows the special rephasing times, \(t^* = M T_{\mathrm{mod}}\), where the boost cavity becomes close to a Fock state, \(\rho_{\mathrm{b}}(t^*) \approx \ket{n_{\mathrm{b}}}\!\bra{n_{\mathrm{b}}}\). One such time is indicated at \(M = 12\) by an arrow. The unstabilized protocol~(a) shows higher occupation at low photon numbers than the stabilized protocol~(b), which results in a standard deviation of \(n_{\mathrm{b}}\) which is more than a factor of two larger in (a) than in (b) at \(t^* = 12 T_{\mathrm{mod}}\). 
    \emph{Parameters:} \(B_0/g_{\mathrm{b}} = 20\), \(\omega_{\mathrm{mod}}/g_{\mathrm{b}} = 1.5\), \(\Delta_{\mathrm{b}}/\omega_{\mathrm{mod}} = (1+\sqrt{5})/2\), \(\Delta_{\mathrm{s}}/B_0 = 1\), initial state \(\ket{g, n_{\mathrm{b}} = 10, n_{\mathrm{s}}=0}\); (b)~\(g_{\mathrm{s}}/g_{\mathrm{b}} = 0\); (c)~\(g_{\mathrm{s}}/g_{\mathrm{b}} = 1\), photon loss from the lossy cavity with rate \(\kappa_{\mathrm{s}}/g_{\mathrm{b}} = 1\) is modeled with a jump operator \(a_{\mathrm{s}}\).}
    \label{fig:boostingzoom}
\end{figure}

The open cavity-qubit system is modelled within the Floquet Lindblad master equation framework, with jump operators accounting for photon loss and qubit relaxation. Specifically, the equation of motion for the joint cavity-qubit density matrix is
\begin{equation}
\frac{d\rho}{dt} = \mathcal{L}_t[\rho] 
= -\frac{i}{\hbar} [H_B^+(t), \rho] + \sum_{j=1}^2 L_j \rho L_j^\dagger - \frac{1}{2} \{L_j^\dagger L_j, \rho \},
\end{equation}
with $L_1=\sqrt{\kappa} a$ (the jump operator modelling photon loss), $L_2 = \sqrt{\Gamma} \sigma^- $ (the jump operator modelling qubit loss, or $T_1$ relaxation), and \(\{A,B\} = AB + BA\) being the anticommutator. In Sec.~\ref{fig:boosting} we use a similar Lindblad framework in a Hilbert space with two cavities, but with \(H_B^+\) replaced by \(H_{\mathrm{boost}}\) and a single jump operator \(L_1 = \sqrt{\kappa_{\mathrm{s}}} a_{\mathrm{s}}\). We solve for \(\rho(t)\) given some initial conditions by direct integration in a truncated cavity Hilbert space (we have checked that the state remains far from the truncation throughout the evolution), using the master equation methods of QuTiP~\cite{qutip}. The data displayed in Figs.~\ref{fig:boosting}, \ref{fig:boostingzoom}, and \ref{fig:theory_exp_comp} is obtained in this way.

\begin{figure}[tb]
    \centering
    \includegraphics[width=0.85\linewidth]{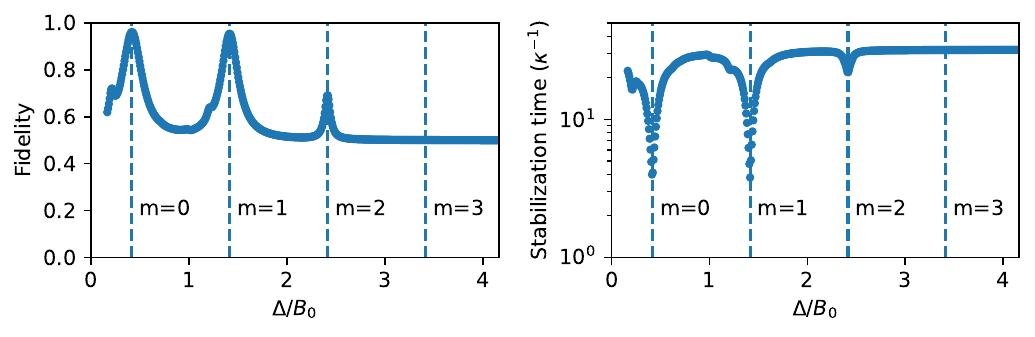}
    \caption{\textbf{Detuning scan at fixed $\omega_\mathrm{mod}/B_0=1$.} 
    Data from Fig.~\ref{fig:nonadiabatictheory_fidelityandlockingtime} for a vertical line cut. The fidelity (time to approach to the steady state) is locally maximum (minimum) when $\Delta  = m \omega_{\mathrm{mod}} + \delta \epsilon$ for $m=0,1,2$. Stabilization is weak at $m=2$, and disappears at $m=3$ because the stabilization rate is smaller than $\Gamma$. \emph{Parameters:} $\omega_\mathrm{mod}/B_0=1$, $g/B_0 = 0.05$, $\kappa/B_0 = 0.05$ and $\Gamma = \kappa/20$.} 
\label{fig:detuningscan_theory}
\end{figure}

Figs.~\ref{fig:nonadiabatictheory}, \ref{fig:nonadiabatictheory_fidelityandlockingtime} and~\ref{fig:detuningscan_theory} display features of the driven-dissipative steady state and the approach to it. The steady state is obtained by numerically finding \(\rho_\alpha(T_{\mathrm{mod}})\) for every member of a basis of initial density matrices \(\rho_\alpha(0)\) (labelled by \(\alpha\)), and hence constructing a matrix for the superoperator $\mathcal{T} \{e^{\int_0^{T_\mathrm{mod}} \mathcal{L}_t \mathrm{d}t }\}$ in this basis. This is the Floquet superoperator which maps an input density matrix to its time evolution after one period. Diagonalizing this superoperator matrix provides the full spectrum of Floquet modes, and in particular the steady state is identified as the mode with decay rate \(0\). The behavior of the steady state within a period is found by again integrating the master equation, now using the steady state of the Floquet superoperator as an initial state.

The diagonalization also provides the inverse timescale on which any initial state of the qubit-cavity system relaxes to the steady state, identified as the stabilization rate. This stabilization rate is of the order of $\kappa$ when the resonance conditions are met, and is otherwise on the scale of $\Gamma$, which is significantly smaller (Figs.~\ref{fig:nonadiabatictheory_fidelityandlockingtime}, ~\ref{fig:detuningscan_theory}). As the higher order resonances have smaller stabilization rates, stabilization vanishes for $m\geq 3$, as Fig.~\ref{fig:detuningscan_theory} shows.

Figures~\ref{fig:boosting} and \ref{fig:boostingzoom} show simulations of the cavity state boosting protocol obtained from numerically integrating the Lindblad master equation for the total system density matrix. This includes the qubit and two cavities (the boost cavity and stabilizing cavity). The distribution of cavity occupations in the boost cavity is computed as
\begin{equation}
    P(n_{\mathrm{b}}, t) = \mathrm{Tr}\left[  \rho(t) \ket{n_{\mathrm{b}}}\!\bra{n_{\mathrm{b}}}\right].
\end{equation}
In the Sec~\ref{sec:pumping}, it is claimed that the stabilized protocol resulted in a boosted state with lower variance. This is visible in Fig.~\ref{fig:boostingzoom}, which shows the same data on a larger scale. At, for instance, \(t^* = 12 T_{\mathrm{mod}}\), the unstabilized protocol produces a state with modal value \(n_{\mathrm{b}}=22\), as does the stabilized protocol. However, the unstabilized state has a standard deviation in \(n_{\mathrm{b}}\) at this time of \(\mathrm{Std}[n_{\mathrm{b}}] \approx 5.98\)---more than twice the value of the stabilized state, which has \(\mathrm{Std}[n_{\mathrm{b}}] \approx 2.72\). The latter is less than the standard deviation of a coherent state with a mean of 22 photons, \(\sqrt{22} \approx 4.69\), so that the stabilized state can outperform a coherent state of the same size in phase estimation measurements~\cite{Giovannetti2011:metrology,Toth2014:metrology_info}.

\section{Onset of stabilization in fast ramp-out measurements}\label{appendix:detuning_sweep_bare_basis}
\begin{figure}[t]
\centering
		\includegraphics[width=0.95\linewidth]{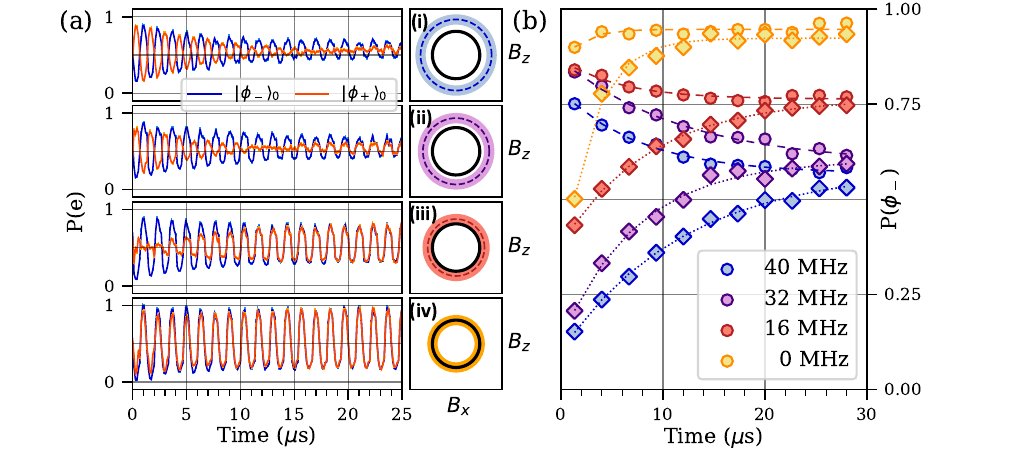}
    \vspace{-0.2cm}
	\caption{\label{fig:detuningscan} 
    \textbf{Onset of stabilization.} 
    (a) The probability of being in state $\ket{e}$ as a function of time for different value of minimum detuning to cavity resonance $\delta/2\pi= (\Delta - B_0)/2\pi$ measured using the fast ramp out protocol. 
    The orange (blue) curves are obtained by starting with the qubit aligned (anti-aligned) with the field, which we denote as $\ket{\phi_+}_0$ and $\ket{\phi_-}_0$, respectively.
    (a.\textit{i}) Far from resonance ($\delta \gg g$), both state initializations damp out to a mixed state while maintaining a $\pi$-phase shift. (a.\textit{ii}-a.\textit{iii}) As the detuning $\delta$ decreases, the state initially prepared in $\ket{\phi_+}$ damps out. Its signal then grows to a non-zero oscillatory amplitude in phase with that from the $\ket{\phi_-}$ state. (a.\textit{iv}) On resonance, both states are quickly stabilized to $\ket{\phi_-(t)}$. (b) Corresponding data acquired using the slow ramp out protocol, effectively showing the envelope of the oscillatory data in panels (a.\textit{i}-a.\textit{iv}). Note that near resonance, the stabilization rate exceeds the oscillation rate leading to significant amplitude variation within a single period, and the slow ramp-out data no longer has a simple correspondence with the envelope of (a).
    Curves are labeled by the minimum detuning to cavity resonance $\delta /2 \pi$ in MHz. Exponential fits to the data are shown in dashed (dotted) lines for preparation in $\ket{\phi_-}$ ($\ket{\phi_+}$). Insets on the right of (a) show a schematic representation of the resonance condition in the $B_x-B_z$ plane. The solid black line denotes the magnetic field strength and the dashed line denotes the mean qubit-cavity detuning (the colored bands show the $\pm g$ width from the cavity-qubit coupling). For both panels, the error bars are smaller than the markers, see appendix \ref{appendix:error_bars} for details.
    }
\end{figure}
In addition to the slow ramp-out data presented in Fig. \ref{fig:adiabatic_detuning_scan}, it is also possible to observe the onset of stabilization in the bare qubit $\ket{g}$, $\ket{e}$ basis using the fast ramp out protocol. 

This protocol gives direct access to the time dependence of the instantaneous eigenstates. States prepared in $\ket{\phi_-}$ (using an initial $\pi$-pulse before the protocol starts) evolve with a $\pi$-phase shift relative to states initialized in $\ket{\phi_+}$ as shown in the blue and orange traces in panels (a.\textit{i})-(a.\textit{ii}) in Fig.~\ref{fig:detuningscan} respectively. As $\delta=\Delta-B_0$ approaches zero, states initially prepared in $\ket{\phi_+}$ damp out before being stabilized in the $\ket{\phi_-}$ state, oscillating at high contrast indefinitely. The instantaneous eigenstate data [shown again in panel (b) for convenience], represents the $\ket{\phi_-}$ population for the oscillatory data presented in panels 
(a.\textit{i})-(a.\textit{iv})
with 0.5 indicating a perfectly mixed state (i.e. zero oscillation contrast). 

The pinch-point, where the oscillation contrast vanishes, observed in the oscillatory data for states prepared initially in $\ket{\phi_+}$ represents the time for the system to reach $P(\phi_-)=0.5$, not the stabilization time. Due to the adiabatic state preparation time before the evolution under $H_B^+$ begins and finite ramp out time,
configurations near the resonance condition [as shown in panel (a.\textit{iv}) and in the $\delta=0$ detuning trace in the adiabatic data in panel (b)] are already mostly in $\ket{\phi_-}$ by the time they are measured even for short evolution times. This is more easily seen in the instantaneous eigenstate data [panel (b)], where the initial value of $P(\phi_-)$ is above $0.5$, before quickly growing to roughly 0.9 after a few microseconds. 

\begin{figure}
    \centering
    \includegraphics[width=0.5\linewidth]{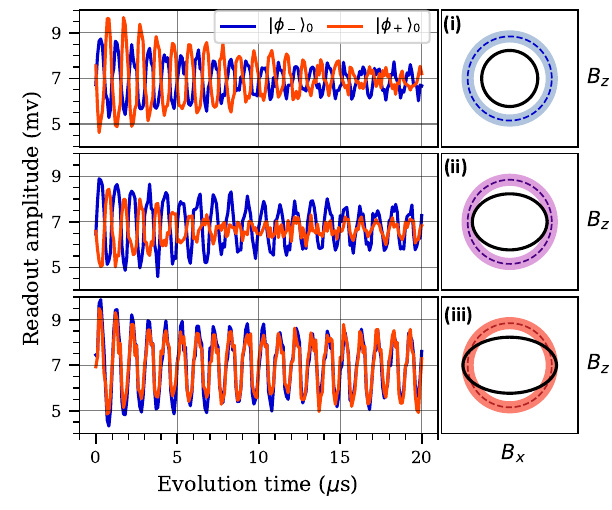}
    \caption{\textbf{Stabilization with elliptical fields} Left: Unnormalized readout amplitude (where a high voltage is proportional to the population in $\ket{g}$ and a low voltage to that in $\ket{e}$) as a function of time for different values of the drive field amplitude $B_x$, for fixed $B_z$ (60 MHz). Blue (orange) traces indicate that the qubit is initialized in $\ket{\phi_-}_0$ ($\ket{\phi_+}_0$), corresponding to the anti-aligned (aligned) state with respect to the applied field. As the amplitude of $B_x$ grows, the quasi-energy states cross resonance with the lossy cavity when $\vec{B}$ points predominantly along $x$, leading to stabilization of the $\ket{\phi_-}$ state. Right: Schematic representation of the applied field (solid black line) and cavity resonance condition (colored dashed line) for different $B_x$ strengths (60, 80, 100 MHz) at a fixed cavity detuning $\Delta=90$~MHz.}
    \label{fig:elliptical_fields}
\end{figure}

Finally, we also observe the stabilization effect for \emph{elliptical} fields where the resonance condition is only satisfied for a portion of the driving period, as shown in Fig. \ref{fig:elliptical_fields}. Here, instead of sweeping the magnitude of $\vec{B}$ to change the detuning, we only sweep the amplitude of the $x$-component of the field, leading to a time-dependent effective detuning between the adiabatic states and the cavity. Just as in the circularly polarized case, when the effective detuning approaches zero, the system is driven to the $\ket{\phi_-}$ state regardless of initial condition and is indefinitely stabilized.

\section{Stabilization Rate and Fidelity}\label{appendix:timescales_and_contrast}

\begin{figure}
    \centering
    \includegraphics[width=\linewidth]{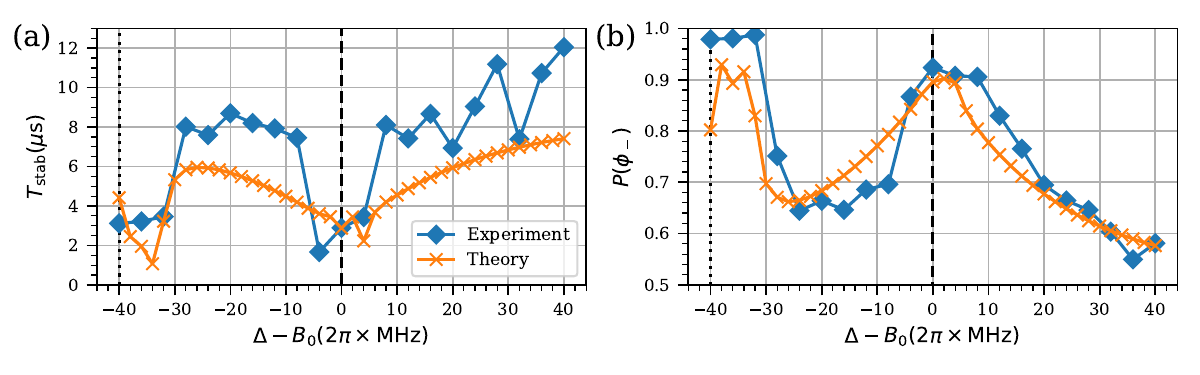}
    \caption{\textbf{Steady state fidelity and stabilization time in the adiabatic regime.} Extracted stabilization time and steady-state $P(\phi_-)$ as a function of the difference between the detuning $\Delta$ of the cavity and the qubit splitting $B_0$. Blue curves (diamond markers) in panels (a) and (b) show the fit results from an exponential fit to the slow ramp out data, including that shown in Fig.~\ref{fig:adiabatic_detuning_scan} in the main text. On resonance ($\Delta-B_0 = 0$, dashed line), the stabilization time $T_{\mathrm{stab}}$ reaches a minimum with corresponding high steady-state $P(\phi_-)$ value. The additional feature at $\Delta-B_0=-B_0/2$ (dotted line) arises due to a two-photon resonance between $\ket{\phi_-,2}$ and $\ket{\phi_+,0}$. Orange curves (x markers) in panels (a) and (b) show the extracted parameters from exponential fits to the Lindblad master equation simulation, which exhibit the same trends as the experimental data.  
    } 
    \label{fig:theory_exp_comp}
\end{figure}

In this section, we compare theory with experiment. Specifically, we extract (i) the steady state value of $P(\phi_-)$, and (ii) the stabilization time constant $T_{\mathrm{stab}}$ 
controlling the decay of $P(\phi_+)$, as a function of the detuning $\delta$. Experimental values for both (i) and (ii) are extracted from an exponential fit to the slow ramp out data. For comparison, the dotted lines in Fig.~\ref{fig:detuningscan}(b) show the exponential fits for $\delta/2\pi = 0,16,32,40$~MHz. 
Panels (a) and (b) of Fig.~\ref{fig:theory_exp_comp} show the extracted time constants and steady state $P(\phi_-)$ for detunings spanning in the range of $\pm 2\pi \times 40$~MHz.

The theoretical curves come from a simulation of the Lindblad master equation~\cite{qutip} described in Appendix~\ref{appendix:numerics_details}, with jump operators for qubit depolarization, qubit dephasing, and cavity decay. The Hamiltonian parameters of the numerical simulation are set to the experimentally calibrated energy-scales shown in Table~\ref{tab:qubit_params}.
The strengths of the cavity and qubit decay jump operators are set to $\kappa = 1/T_\mathrm{cav}$ and $\Gamma_q = 1/T_1$, using the experimentally measured values shown in Appendix~\ref{appendix:qubit_timescales}. 

An appropriate treatment of pure dephasing is more challenging because the Lindblad master equation fundamentally relies on a Markov approximation, and, as a result, it only fully captures dephasing in the presence of white noise. The noise spectrum of qubit-frequency fluctuations present in our devices is not white, as evidenced by the Gaussian decay of coherence in Ramsey measurements, shown in Fig.~\ref{fig:qubit_timescales}. 
Setting the dephasing rate to $1/T_2$, with $T_2$ obtained from Ramsey measurements, vastly overestimates dephasing in the presence of the strong drive.
We therefore approximate the experiment using an effective pure dephasing rate $1/T_d$, where $T_d =10.2\ \mu$s is the drive following time shown in Fig.~\ref{fig:hourglass}(a), since this rate more closely captures the effect of the experimental noise spectrum on adiabatic following of the drive field.

Both the experimental and theory traces show a clear feature on resonance ($\Delta-B_0=0$) with a peak in the steady state $P(\phi_-)$ accompanied by a steep drop in the time required to reach the steady state, consistent with the minimum stabilization time of roughly $2/\kappa$ predicted by Floquet Lindblad calculation in Appendix \ref{appendix:numerics_details}.
In addition to the single photon resonance between $\ket{\phi_+,0}$ and $\ket{\phi_-,1}$, both theory and experiment show and additional two-photon resonance at $\delta = -B_0/2$, where $\ket{\phi_-,2}$ and $\ket{\phi_+,0}$ become resonant. This two-photon resonance is a generic feature and persists even outside the adiabatic regime explored in the experiment. Indeed, peaks in the steady state $P(\phi_-)$ associated with two-photon resonances are visible in Fig.~\ref{fig:nonadiabatictheory} even at high frequency (indicated by dotted lines).

The theory shows good qualitative agreement with the experimental results. Discrepancies in the exact values of the stabilization time and steady-state $P(\phi_-)$ are likely due to the inability of the Lindblad treatment to capture the colored noise present in our experiment. In principle, if the noise power spectrum of fluctuations in the qubit frequency is known, alternative simulation techniques such as the quantum trajectories method can be used to exactly model the effect of the colored noise \cite{Carmichael:1993}. Instead of computing the dynamics of the density matrix, as is done in the Lindblad master equation formalism, these methods compute the exact time dynamics of the system for a particular noise realization, which are then averaged over simulated instances of the noise. The noise spectrum for our device is not sufficiently well-characterized to permit first-principle modeling along these lines. However, our experimental results show that the effect of $T_2$ processes are strongly suppressed, to the point that we obtain qualitative agreement using the Lindblad master equation formalism with $T_1$ processes alone. In addition, as we showed in related work \cite{syntheticFields}, the system decoherence rate $T_d$ is independent of the driving field strength $B_0$ and rotation rate $\omega_{\mathrm{mod}}$ for a wide range of drive parameters, indicating that the leading order dynamics is well characterized by our simple model.

\section{Qubit characterization}
\label{appendix:qubit_timescales}

\begin{figure*}
    \centering
    \includegraphics[width=0.9\linewidth]{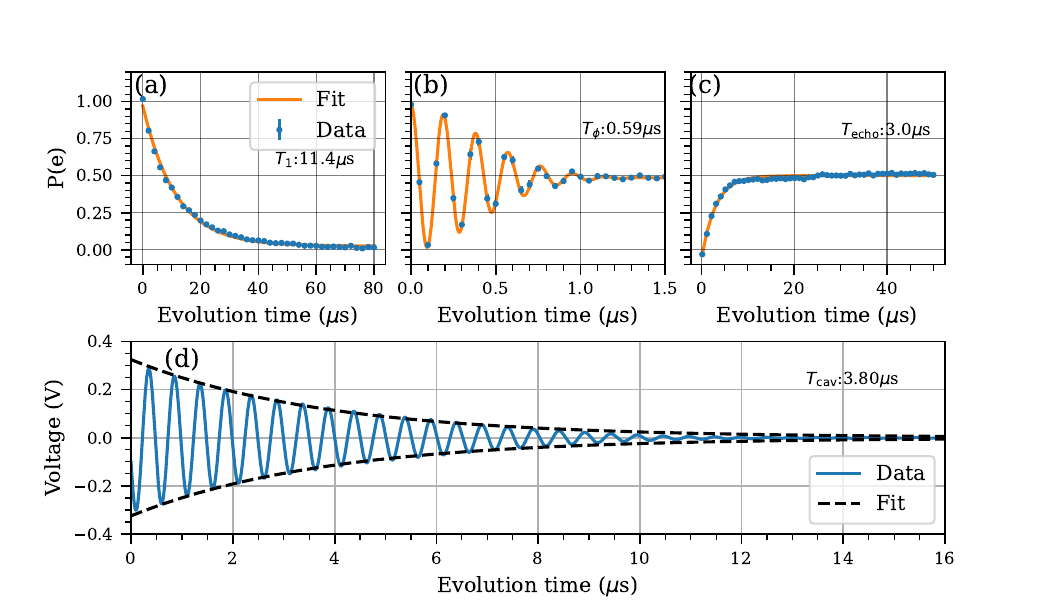}
    \caption{\textbf{System coherence times.} (a) $\ket{e}$ state population as a function of delay time after qubit excitation. An exponential fit gives $T_1=11.4\ \mu$s.  (b) Ramsey oscillations using a $5$~MHz virtual detuning on the second $\pi/2$-pulse to extract $T_{\phi}=590$~ns.  (c) $\ket{e}$-state population after a standard Hahn echo sequence, giving $T_{\mathrm{echo}}=3\ \mu$s. The six-fold increase between $T_{\phi}$ and $T_{\mathrm{echo}}$ indicates the presence of a large low-frequency noise component in the qubit frequency. (d) Heterodyne measurement of the main cavity field after populating it with a coherent state. The field decay time $T_{\mathrm{cav}} = 3.8\ \mu$s is extracted from a fit to a damped oscillatory response (envelope of the fit is indicated by dashed lines). The photon lifetime corresponds to $T_{\mathrm{cav}}/2$. Presented data is the average of $10$ repeated traces, each consisting of $10^4$ shots per time point. Error bars are computed as the standard error of the mean of the repeated traces. The error bars are smaller than the marker size for panels (a), (b), (c) and not shown for panel (d). }
    \label{fig:qubit_timescales}
\end{figure*}

We perform standard characterization of the various decoherence timescales in the qubit-cavity system with the qubit at $4.9$~GHz, far detuned from the main cavity. Corresponding measurements are shown in Fig. \ref{fig:qubit_timescales}.
We extract the qubit decay time $T_1 = 11.4\ \mu$s from an exponential fit to the data after initializing the qubit in the $\ket{e}$ and waiting a time $\tau$, with the qubit far detuned from both main cavity. Purcell loss from the main cavity provides a negligible contribution to measurement, so $T_1 = 11.4\ \mu$s, therefore, represents the decay rate of the qubit into channels other than the main cavity. 

To find pure dephasing time $T_{\phi} = 590$~ns, we perform a Ramsey-style measurement where two $\pi/2$ pulses are applied to the qubit with a variable wait time. However, to maintain a large separation of scales between the oscillatory and decaying time constants of the Ramsey signal, we rotate the phase of the second pulse at $5$~MHz to apply a virtual detuning between the drive and qubit without loss of contrast. We use a gaussian envelope to compute the dephasing time~\cite{Krantz:2019} instead of the regular exponential envelope, due to the low-frequency noise present in tunable frequency qubits away from a sweet spot. We extract $T_{\mathrm{echo}} = 3\ \mu$s from the standard Hahn echo sequence and fit an exponential model to the data. 

Finally, the cavity ringdown time, $T_{\mathrm{cav}} =3.8\ \mu$s, is measured by populating the cavity with a coherent state and fitting the resulting leakage signal after turning off the cavity drive. Since heterodyne detection measures the cavity field \emph{amplitude}, we fit to a damped sinusoidal response. The photon lifetime in the cavity is \emph{half} the ringdown time as it scales with the power in the cavity instead of the field amplitude giving a final lifetime of $1.9\ \mu$s. 

\section{Experimental error bars and qubit state readout}\label{appendix:error_bars}

The qubit state is read out using standard two-tone pulsed spectroscopy on the readout cavity~\cite{Blais_2021, Krantz:2019}. In this method, $\ket{g}$ and $\ket{e}$ correspond to two different resonant frequencies of the readout cavity. The qubit state can be inferred by measuring the amplitude and phase of a microwave pulse at the cavity frequency, which can be converted to IQ points (the cartesian representation of the sine and cosine components of the cavity response). By taking reference $\ket{g}$ and $\ket{e}$ measurements, we can convert an output voltage to a normalized qubit population:
\begin{equation}
    P(e) = \frac{\overrightarrow{IQ}_{\mathrm{data-g}}\cdot\overrightarrow{IQ}_{e-g}}{|\overrightarrow{IQ}_{e-g}|^2},
\end{equation}
where $\overrightarrow{IQ}_{\mathrm{data-g}}$ is vector-valued difference between the resonator response for the given configuration and the response for $\ket{g}$. 
The quantity $\overrightarrow{IQ}_{\mathrm{e-g}}$ is the analogous quantity for a reference measurement of a qubit in the $\ket{e}$ state.
For all data in the text, the reference $\ket{e}$ trace was acquired at the beginning of a measurement run (i.e. for a given detuning) by averaging $5\times 10^4$ single shots, and the $\ket{g}$ reference trace was acquired every $10$ measurement points as the average of $10^4$ shots. 

Data points in the main text was acquired by averaging $10^4$ shots at a $5$~kHz repetition rate to avoid spurious $\ket{e}$ population between experimental runs. Due to different readout amplitudes for $\ket{g}$ and $\ket{e}$ outcomes, the measurement uncertainties are not constant and need to be characterized as a function $P(e)$. 
We estimate the typical readout error for a given $\ket{g}$-$\ket{e}$ superposition by interpolating from a set of reference uncertainty measurements. These reference uncertainties are computed by taking the standard deviation of $10$ repeated measurements of one period of oscillation. All the data presented in the main text uses this method as a typical readout uncertainty instead of computing an error bar within each dataset.

The qubit characterization data, shown in Fig.~\ref{fig:qubit_timescales}, is the average of $10$ repeated measurements, each consisting of the average of $10^4$ single shots (which are not individually recorded). The uncertainty for these measurements is computed as the standard error of the mean 
($\mathrm{Std}/\sqrt{N-1}$ with \(N=10\))
and is typically smaller than the marker size. 

\end{document}